\documentclass[
	aps,
	amsmath,amssymb,
	nofootinbib,
	twocolumn,
	notitlepage,
	superscriptaddress
   ]{revtex4-1}

\newcommand{\Tr}{\text{Tr}}
\usepackage{graphicx}
\usepackage{color}
\usepackage{amsmath}
\usepackage{dcolumn}
\usepackage{siunitx}
\usepackage{bm}
\usepackage{braket}
\usepackage{here}

\makeatletter
\let\MYcaption\@makecaption
\makeatother

\usepackage{subcaption}
\makeatletter
\let\@makecaption\MYcaption
\makeatother

\begin{document}


\title[]{Experimental quantum kernel machine learning with nuclear spins in a solid}

\author{Takeru Kusumoto$^+$}
\affiliation{Graduate School of Engineering Science, Osaka University, 1-3 Machikaneyama, Toyonaka, Osaka 560-8531, Japan.}

\author{Kosuke Mitarai$^+$}
\email{mitarai@qc.ee.es.osaka-u.ac.jp}
\affiliation{Graduate School of Engineering Science, Osaka University, 1-3 Machikaneyama, Toyonaka, Osaka 560-8531, Japan.}
\affiliation{QunaSys Inc., High-tech Hongo Building 1F, 5-25-18 Hongo, Bunkyo, Tokyo 113-0033, Japan}

\author{Keisuke Fujii}
\affiliation{Graduate School of Engineering Science, Osaka University, 1-3 Machikaneyama, Toyonaka, Osaka 560-8531, Japan.}
\affiliation{Quantum Information and Quantum Biology division, Institute for Open and Transdisciplinary Research Initiatives, Osaka University, Japan}
\affiliation{JST, PRESTO, 4-1-8 Honcho, Kawaguchi, Saitama 332-0012, Japan}

\author{Masahiro Kitagawa}
\affiliation{Graduate School of Engineering Science, Osaka University, 1-3 Machikaneyama, Toyonaka, Osaka 560-8531, Japan.}
\affiliation{Quantum Information and Quantum Biology division, Institute for Open and Transdisciplinary Research Initiatives, Osaka University, Japan}

\author{Makoto Negoro}
\email{negoro@qc.ee.es.osaka-u.ac.jp}
\affiliation{Quantum Information and Quantum Biology division, Institute for Open and Transdisciplinary Research Initiatives, Osaka University, Japan}
\affiliation{JST, PRESTO, 4-1-8 Honcho, Kawaguchi, Saitama 332-0012, Japan}

\date{\today}

\begin{abstract}
	We employ so-called quantum kernel estimation to exploit complex quantum dynamics of solid-state nuclear magnetic resonance for machine learning.
We propose to map an input to a feature space by input-dependent Hamiltonian evolution, and the kernel is estimated by the interference of the evolution.
Simple machine learning tasks, namely one-dimensional regression tasks and two-dimensional classification tasks, are performed using proton spins which exhibit correlation over 10 spins.
We also performed numerical simulations to evaluate the performance without the noise inevitable in the actual experiments.
The performance of the trained model tends to increase with the longer evolution time, or equivalently, with a larger number of spins involved in the dynamics for certain tasks.
This work presents a quantum machine learning experiment using one of the largest quantum systems to date.
\end{abstract}

\pacs{Valid PACS appear here}
\maketitle

\def\thefootnote{+}\footnotetext{These authors contributed equally to this work}\def\thefootnote{\arabic{footnote}}

\section{Introduction}
Quantum machine learning is an emerging field that has attracted much attention recently.
The major algorithmic breakthrough was an algorithm invented by Harrow-Hassidim-Lloyd \cite{Harrow2009}.
This algorithm has been further developed to more sophisticated machine learning algorithms \cite{Biamonte2017, mariaschuld2018}.
However, a quantum computer that is capable of executing those algorithms is yet to be realized.
At present, noisy intermediate-scale quantum (NISQ) devices \cite{Preskill2018}, which consist of several tens or hundreds of noisy qubits, are the most advanced technology.
Although their performance is limited compared to the fault-tolerant quantum computer, simulation of the NISQ devices with 100 qubits and sufficiently high gate fidelity are beyond the reach for the existing supercomputer and classical simulation algorithms \cite{Boixo2018, Villalonga2019, Arute2019}.
This fact motivates us to explore its power for solving practical problems.

Many NISQ algorithms for machine learning have been proposed in recent works \cite{Mitarai2018bm, Schuld2019, Benedetti2019, Dallaire-Demers2018, Cong2019, Liu2018, Huggins2019, Havlicek2019, Fujii2017, Ghosh2019}.
Almost all of the algorithms require us to evaluate an expectation value of an observable, which is sometimes troublesome to measure by sampling, for example with superconducting or trapped-ion qubits.
On the other hand, NMR can evaluate the expectation value with a one-shot experiment owing to its use of a vast number of duplicate quantum systems.
It is therefore, in fact, a great testbed for those algorithms.
A major weak point of NMR is that its initialization fidelity is quite low; at the thermal equilibrium of room temperature, the proton spins can effectively describe with a density matrix $\rho_{eq} = \frac{1}{2}(I+\epsilon I_z)$ with $\epsilon \approx 10^{-5}$.
Nevertheless, ensemble spin systems can exhibit complex quantum dynamics that are classically intractable.
For example, the dynamical phase transition between localization and delocalization has been observed in polycrystalline adamantane along with tens of correlated proton spins \cite{Alvarez2015}.
Discrete time-crystalline order has been observed in disordered \cite{Choi2017} and ordered \cite{Rovny2018, Rovny2018a} spin systems.

In this work, we employ such an ensemble spin system for machine learning. 
Specifically, we implement the kernel-based algorithm which utilizes the quantum state as a feature vector and is a variant of theoretical proposals \cite{Mitarai2018bm, Schuld2019, Farhi2018}.
The experimental verification has been provided in Refs.~\cite{Havlicek2019, Bartkiewicz2019} using either superconducting qubits or the photonic system.
Our strategy to use the NMR is advantageous in that we can estimate the value of the kernel, which is the inner product of two quantum states, by single-shot experiments.
We perform simple regression and binary classification tasks using the dynamics of nuclear spins in polycrystalline adamantane sample.
Also, to carry out the performance analysis of our approach without the inevitable effect of noise in experiments, we present numerical simulations of 20 spin dynamics.
For certain tasks, we observed that the performance of the trained model becomes better as more spins are involved in the dynamics.
We employ one of the largest quantum systems to date for a quantum machine learning experiment in this work.

\begin{figure*}
\begin{minipage}[b]{0.48\linewidth}
\centering
\includegraphics[width=0.3\linewidth]{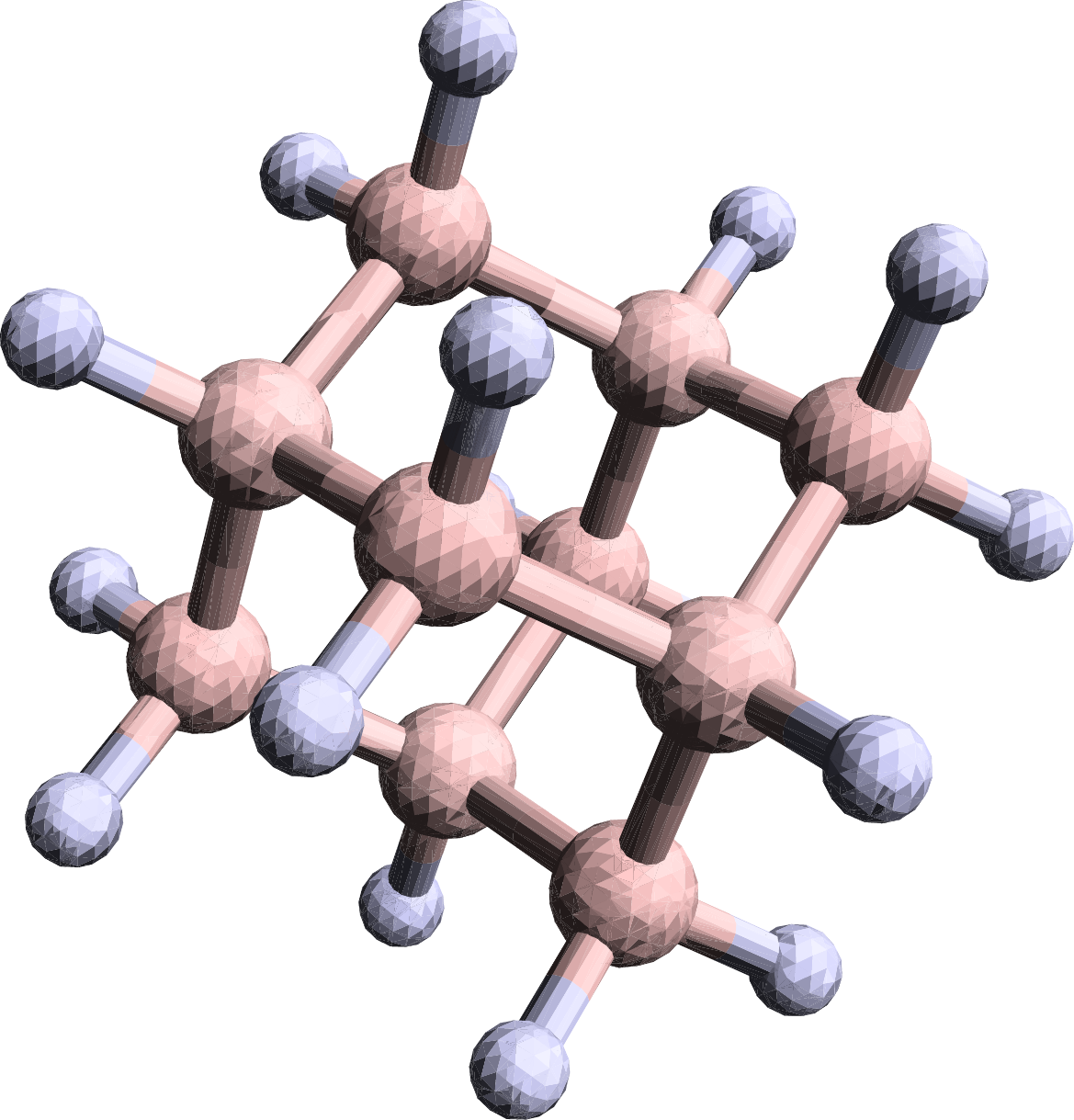}
\subcaption{}
\end{minipage}
\begin{minipage}[b]{0.48\linewidth}
\centering
\includegraphics[width=\linewidth]{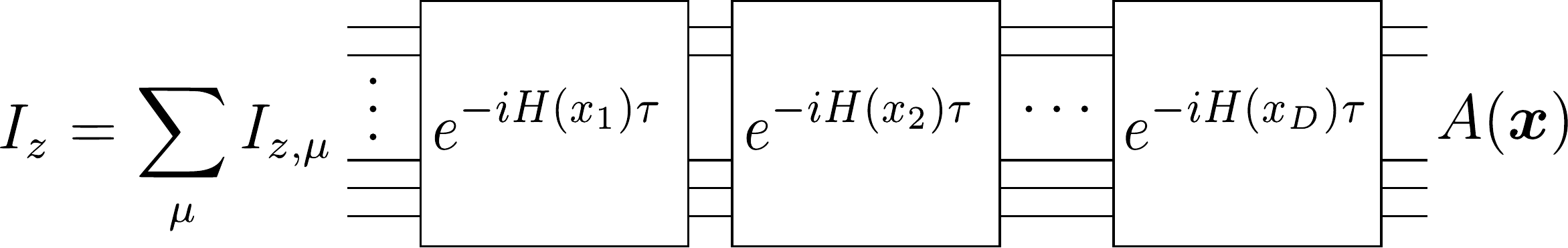}
\subcaption{}
\end{minipage}
\\
\begin{minipage}[b]{0.48\linewidth}
\centering
\includegraphics[width=\linewidth]{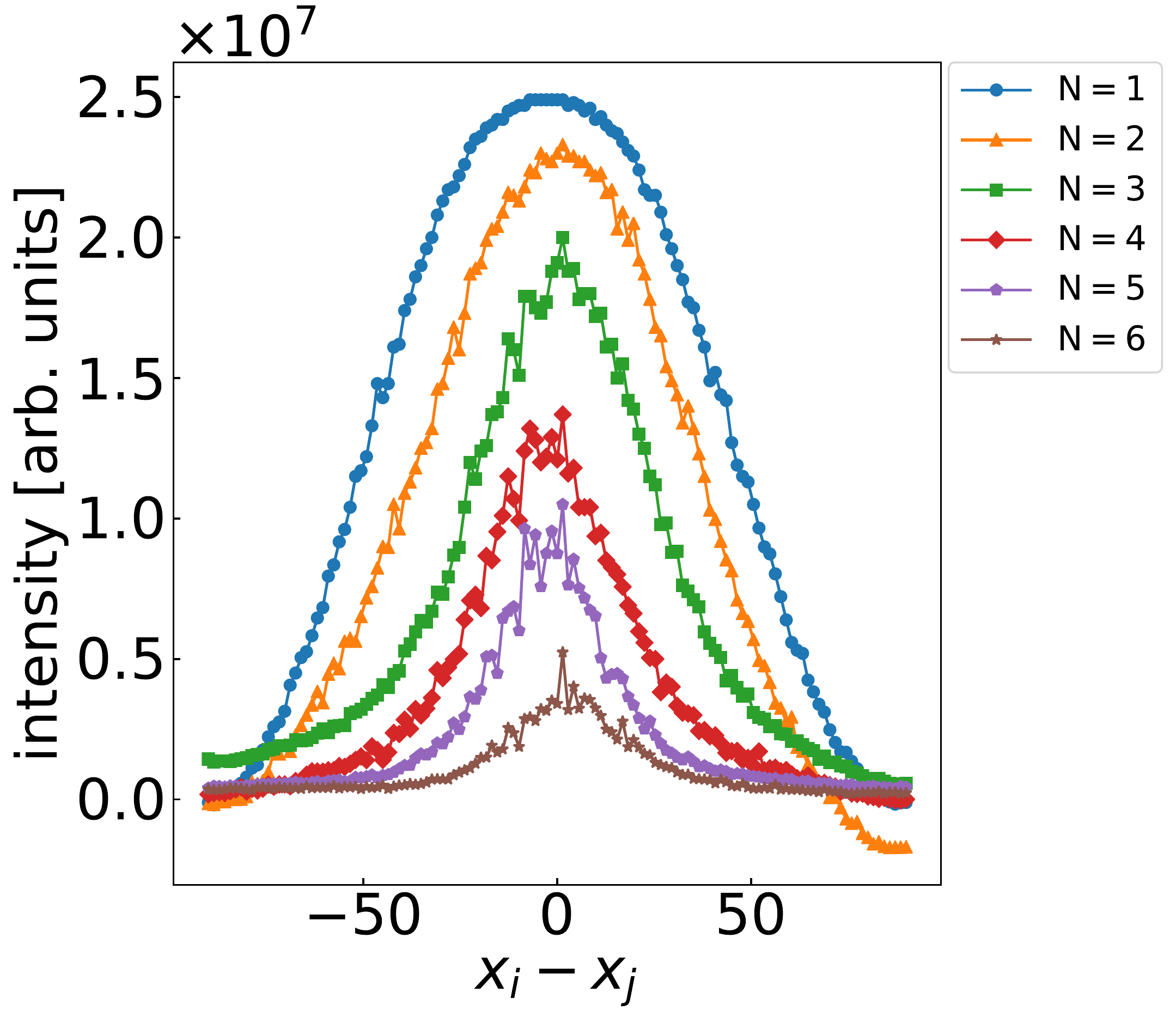}
\subcaption{}
\end{minipage}
\begin{minipage}[b]{0.48\linewidth}
\centering
\includegraphics[width=\linewidth]{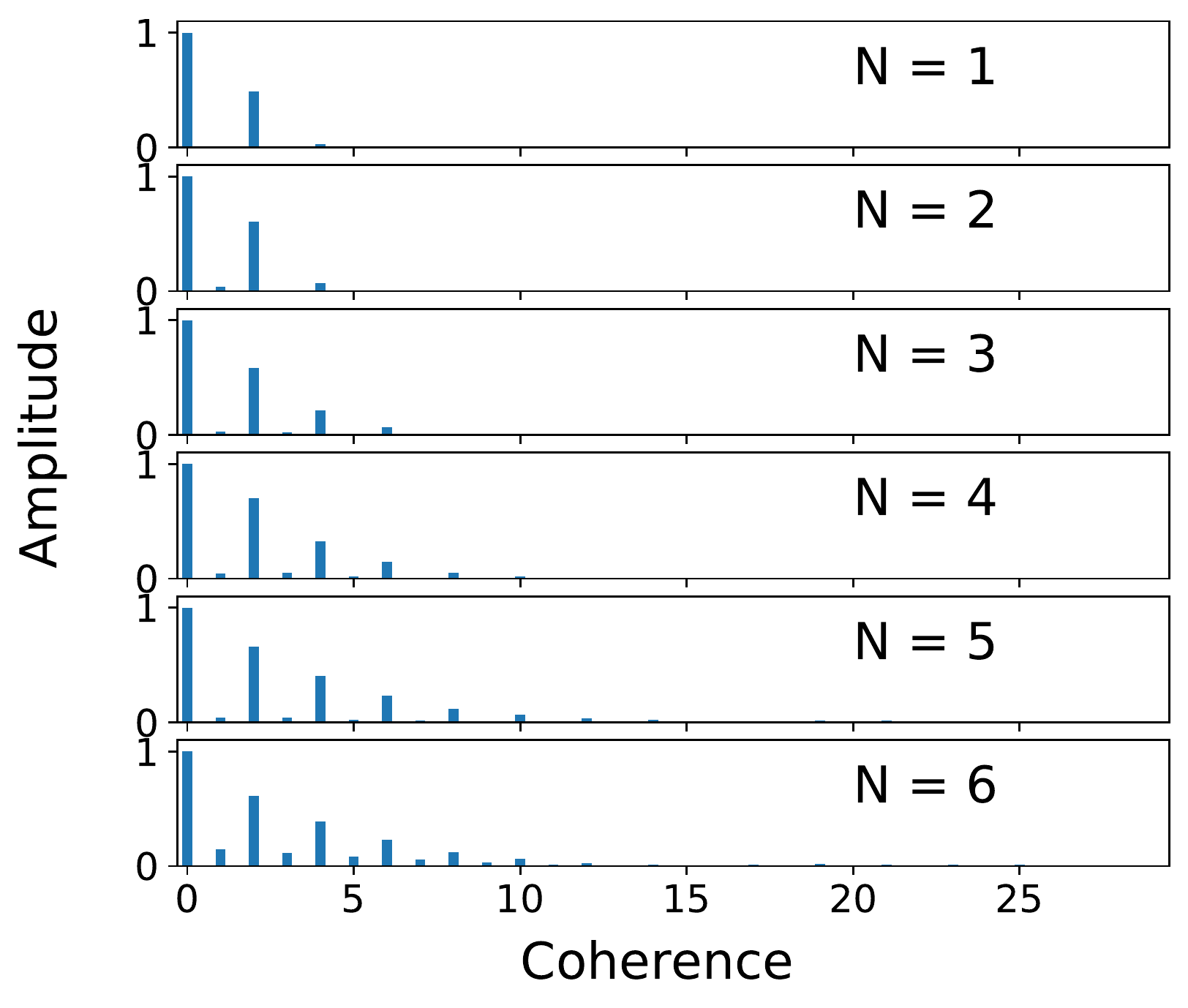}
\subcaption{}
\end{minipage}
\caption{(a) Adamantane molecule. (b) Quantum circuit employed in this work to realize data-dependent quantum state $\rho(\bm{x})$. (c) NMR kernel employed in this work. (d) Fourier transform of (c) which corresponds to the obtained $^1\text{H}$ multiple-quantum spectra for $N = 1$ to $N = 6$.}
\label{fig:1DKernel}
\end{figure*}

\section{Theory}

\subsection{Kernel methods in machine learning}
In machine learning, you are asked to extract some patterns, or features, in a given dataset \cite{Bishop2011, mariaschuld2018}.
It is sometimes useful to pre-process them beforehand to achieve the objective.
For example, a speech recognition task might become easier when we work in the frequency domain; in this case, the useful pre-processing would be Fourier transform.
The space in which such pre-processed data live is called feature space.
For a given set of data $\{\bm{x}_i\}_{i=1}^{N_{d}}\subset \mathbb{R}^D$, a feature space mapping $\bm{\phi}(\bm{x})$ constructs the data in the feature space $\{\bm{\phi}(\bm{x}_i)\}_{i=1}^{N_{d}} \subset \mathbb
{R}^{D_f}$.
The feature map has to be carefully taken as to maximize the performance in e.g. classification tasks.

The kernel methods are a powerful tool in machine learning.
It uses a distance measure of two inputs defined as a kernel, $k:\mathbb{R}^D\times\mathbb{R}^D \to \mathbb{R}$.
For example, a kernel can be defined as an inner product of two feature vectors: 
\begin{align}
    k(\bm{x}_i, \bm{x}_j) = \bm{\phi}(\bm{x}_i)^T\bm{\phi}(\bm{x}_j).
\end{align}
Many machine learning models, such as support vector machine or linear regression, can be constructed using the kernel only, that is, we do not have to explicitly hold $\bm{\phi}(\bm{x})$.
For example, for a given teacher dataset $\{y_i\}_{i=1}^N\subset\mathbb{R}$ each corresponding to the input $\bm{x}_i$, a linear regression model can be constructed by,
\begin{align}\label{eq:kernel_LR}
    f_{\text{LR}}(\bm{x}) &= \bm{y}^TK^{-1}\bm{k}(\bm{x}),
\end{align}
where we have defined the $N_d$-dimensional vector $\bm{k}(\bm{x})$,
\begin{align}
    \bm{k}(\bm{x}) = \left(
        \begin{array}{c}
            k(\bm{x}_1, \bm{x})\\
            k(\bm{x}_2, \bm{x})\\
            \vdots\\
            k(\bm{x}_{N_d}, \bm{x})
        \end{array}
        \right).
\end{align}
and a matrix $K=\{K_{ij}\}_{i,j=1}^{N_d}=\{k(\bm{x}_i, \bm{x}_j)\}_{i,j=1}^{N_d}$.
$f_{\mathrm{LR}}$ is equivalent to defining
\begin{equation}
    f_{\text{LR}}(\bm{x}) = \bm{w}^T \bm{\phi}(\bm{x}),
\end{equation}
where $\bm{w}\in \mathbb{R}^{D_f}$ is chosen to minimize the mean squared error between the model prediction and the teacher;  $\bm{w} = \mathrm{argmin}_{\bm{w}}\sum_i|\bm{w}^T \bm{\phi}(\bm{x})-\bm{y}_i|^2$.

\subsection{Implementing kernel by NMR}
In NMR, we can prepare a data-dependent operator $A(\bm{x}_i)$ by applying a data-dependent unitary transformation $U(\bm{x}_i)$ on the initial $z$-magnetization $I_z = \sum_{\mu=1}^n I_{z,\mu}$, that is, $A(\bm{x}_i) = U(\bm{x}_i) I_z U^\dagger(\bm{x}_i)$.
Here, $I_{\alpha, \mu}$ $(\alpha=x,y,z)$ is the $x,y,z$ component of the spin operator of the $\mu$-th spin and $n$ represents the number of spins.
$A(\bm{x}_i)$ with a sufficiently large $n$ is generally intractable by classical computers \cite{Morimae2014, Fujii2018}.
We employ this operator $A(\bm{x}_i)$ as a feature map $\bm{\phi}_{\mathrm{NMR}}(\bm{x}_i)$.
$A(\bm{x}_i)$ can be regarded as a vector, for example, by expanding $A(\bm{x}_i)$ as a sum of Pauli operators.
For an $n$-spin-1/2 system, $A(\bm{x}_i)$ is a vector in $\mathbb{R}^{4^n}$.
The dynamics of NMR can involve tens of spins maintaining its coherence \cite{Krojanski2004, Krojanski2006, Alvarez2010, Alvarez2015}, which means we can employ an approximately $4^{O(10)}$ dimensional feature vector for machine learning.
Although the high dimensional feature space does not always mean the superiority in machine learning tasks, the fact that we can work with the feature space which has been intractable with a classical computer motivates us to explore its power.

The kernel method opens up a way to exploit $A(\bm{x}_i)$ directly for machine learning purposes.
While we cannot evaluate each element of $A(\bm{x}_i)$ because it takes exponential amount of time, we can evaluate the inner product of two feature vector $A(\bm{x}_i)$ and $A(\bm{x}_j)$ efficiently.
To see this, let us define the inner product of two feature vectors $\bm{\phi}_{\mathrm{NMR}}(\bm{x}_i)$ and $\bm{\phi}_{\mathrm{NMR}}(\bm{x}_j)$ as,
\begin{align}
    k_{\mathrm{NMR}}(\bm{x}_i, \bm{x}_j) &= \bm{\phi}_{\mathrm{NMR}}(\bm{x}_i)^T\bm{\phi}_{\mathrm{NMR}}(\bm{x}_j) \\
    &= \Tr \left(A(\bm{x}_i)A(\bm{x}_j)\right).
\end{align}
Then,
\begin{align}
    \Tr \left(A(\bm{x}_i)A(\bm{x}_j)\right) &= \Tr \left(U(\bm{x}_i)I_z U^\dagger(\bm{x}_i)U(\bm{x}_j)I_z U^\dagger(\bm{x}_j)\right)\nonumber \\
    &= \Tr \left(U^\dagger(\bm{x}_j)U(\bm{x}_i)I_zU^\dagger(\bm{x}_i)U(\bm{x}_j)I_z\right).\label{eq:qkernel}
\end{align}
This can easily be evaluated by NMR.
Note that at the thermal equilibrium, the density matrix of spin systems is $\rho_{eq} = \frac{1}{2^n}\left(I+\epsilon I_z + o(\epsilon^2)\right)$.
Assuming $\epsilon \ll 1$, the above inner product can be evaluated from the following quantity,
\begin{align}
    &\bm{\phi}_{\mathrm{NMR}}(\bm{x}_i)^T\bm{\phi}_{\mathrm{NMR}}(\bm{x}_j)\nonumber \\
    &\propto \Tr \left(U^\dagger(\bm{x}_j)U(\bm{x}_i)\rho_{eq}U^\dagger(\bm{x}_i)U(\bm{x}_j)I_z\right),
\end{align}
that is, we first evolve the system with $U(\bm{x}_i)$ and then with $U^\dagger(\bm{x}_j)$, and finally measure $I_z$.
Note that when $\|\bm{x}_i-\bm{x}_j\|\approx 0$, the protocol resembles the famous Loschmidt echo \cite{GORIN200633, Sanchez2009}.
A similar protocol is also used for measuring out-of-time-ordered correlator (OTOC) \cite{Grttner2017,Wei2018}, which is considered as a certain complexity measure of quantum many-body systems.

\section{Experiment}
We propose to use $U(\bm{x})$ for an input $\bm{x}=\{x_j\}_{j=1}^D$ that takes the form of,
\begin{align}\label{eq:uforkernel}
U\left(\bm{x} \right) = e^{-iH(x_D)\tau}\cdots e^{-iH(x_2)\tau}e^{-iH(x_1)\tau},
\end{align}
where $\tau$ is a constant and $H(x_j)$ is an input-dependent Hamiltonian (Fig. \ref{fig:1DKernel} (b)).
In this work, we choose $H(x_j)$ to be
\begin{align}
    H(x_j) = e^{-ix_jI_z}\sum_{\mu<\nu} d_{\mu\nu}\left(I_{y,\mu}I_{y,\nu} - I_{x, \mu}I_{x, \nu}\right)e^{ix_j I_z},
\end{align}
where $I_\alpha=\sum_m I_{\alpha m}$.
The Hamiltonian $H(0)$ can approximately be constructed from the dipole interaction among nuclear spins in solids with a certain pulse sequence \cite{Baum1985, Warren1980, Alvarez2015} described in Supplementary Material with details of the experiment.
Shifting the phase of the pulse by $x$ provides us $H(x)$ for general $x$.
This Hamiltonian with $x_j=0$ created in adamantane has been shown to have a delocalizing feature in Refs. \cite{Krojanski2004, Krojanski2006, Alvarez2010, Alvarez2015}, which makes it appealing as we wish to involve as many spins as possible in the dynamics.

To illustrate character of the kernel function, we show the shape of the kernel for one-dimensional input $x$ obtained with this sequence setting $\tau=N\tau_1$ where $\tau_1 = 60\mathrm{\mu s}$ and $N=1, 2, \cdots, 6$ as Fig. \ref{fig:1DKernel} (c).
Since $H(x)$ is defined with the varying phase of the pulse, the value of the kernel $k(x_i,x_j)$ for two one-dimensional inputs $x_i$ and $x_j$ only depends on their difference, $x_i-x_j$.
We therefore show the value of the kernel as a function of $x_i-x_j$ in Fig. \ref{fig:1DKernel} (c).
The decay of the intensity of the signal with increasing $N$ is due to decoherence. 
We show the Fourier transform of the measured NMR kernel (Fig.~\ref{fig:1DKernel} (c)) in Fig.~\ref{fig:1DKernel} (d).
The frequency component of an integer $m$ in this experiment, which is called coherence order, results from the existence of $m$-body spin operators in $A(\bm{x})$, and its intensity is called multiple quantum spectrum \cite{Grttner2017, Alvarez2010, Alvarez2015}.
Fig.~\ref{fig:1DKernel} (d) indicates that the dynamics involving 10 spins is present for $N=6$ \cite{baum1985multiple}.

\subsection{One-dimensional regression task}
As the first demonstration, we perform the one-dimensional kernel regression task using the kernel shown in Fig, \ref{fig:1DKernel} (c).
To evaluate the nonlinear regression ability of the kernel, we use $y=\sin(2\pi x/50)$ and $y=\frac{\sin (2\pi x/50)}{2\pi x/50}$, which will be refered to as sin and sinc function, respectively.
We randomly drew 40 samples of $x$ from $[-45, 45]$ (in degrees) to construct the traning data set which consists of the input $\{x_j\}_{j=1}^{40}$ and the teacher $\{y_j\}_{j=1}^{40}$ calculated at each $x_j$.
The NMR kernel $k_{\mathrm{NMR}}(x_i,x_j)$ is measured for each pair of data to construct the model by kernel Ridge regression~\cite{Bishop2011}.
We let the model predict $y$ for 64 $x$'s including the training data.
The regularization strength was chosen to minimize the mean squared error of the result at the 64 evaluation data.

The result for the sin function is shown in Fig.~\ref{fig:sin} (a)-(f).
That for the sinc function is shown in Supplementary Material.
Fig. \ref{fig:1DMSE} (a) shows the accuracy of learning evaluated by the mean squared error between the output from the trained model and true function.
We see that the regression accuracy tends to increase with a larger $N$.
However, because of the deteriorating signal-to-noise ratio, the result also gets noisy with increasing $N$.

\begin{figure*}
 \begin{minipage}[b]{0.49\linewidth}
 \begin{minipage}[b]{0.48\linewidth}
  \centering
  \includegraphics[width=37mm]
  {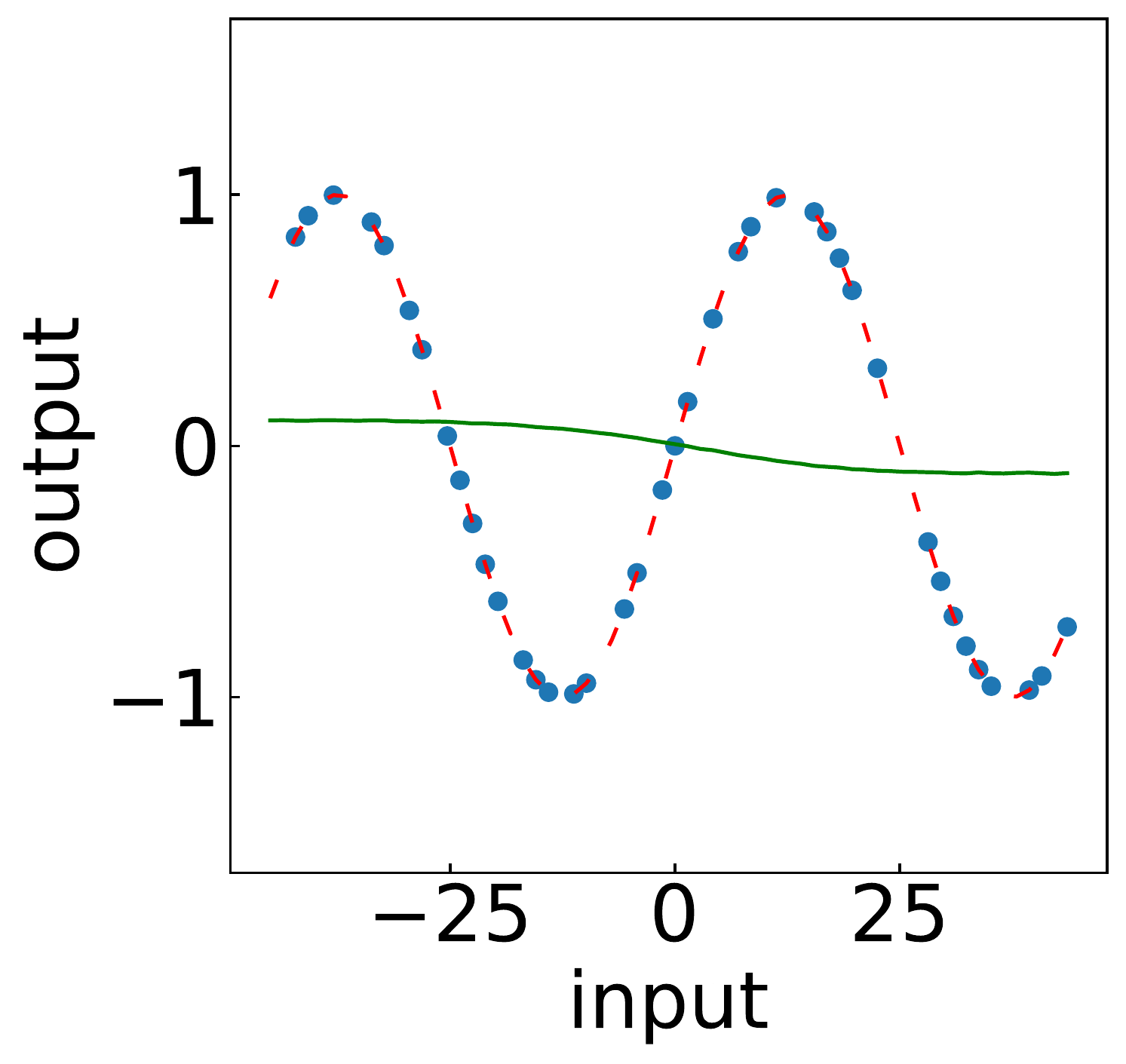}\vspace{-3truemm}
  \subcaption{N = 1}\label{sin1}
 \end{minipage}
 \begin{minipage}[b]{0.48\linewidth}
  \centering
  \includegraphics[width=37mm]
  {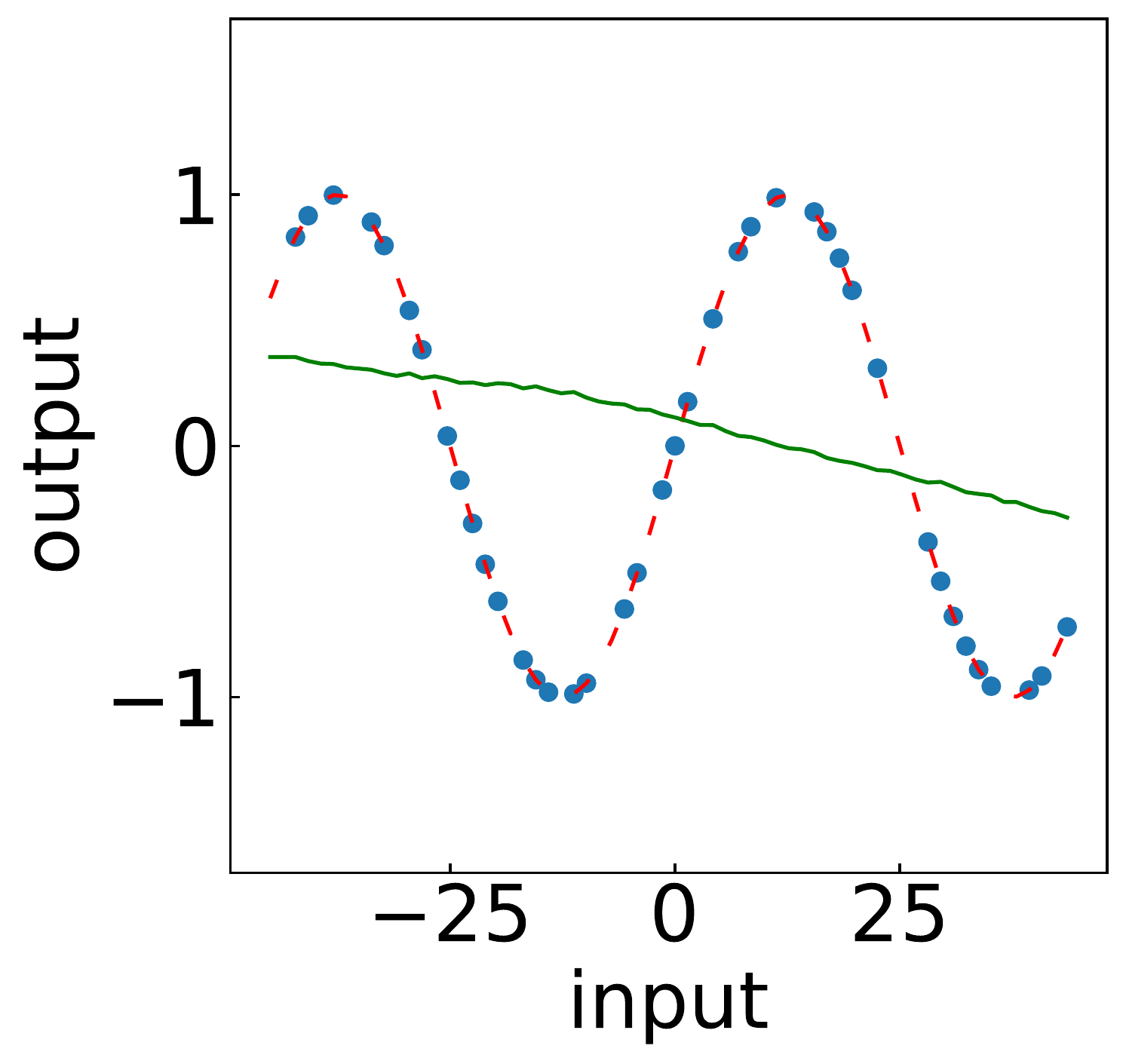}\vspace{-3truemm}
  \subcaption{N = 2}\label{sin2}
 \end{minipage}\\\vspace{1truemm}
 \begin{minipage}[b]{0.48\linewidth}
  \centering
  \includegraphics[width=37mm]
  {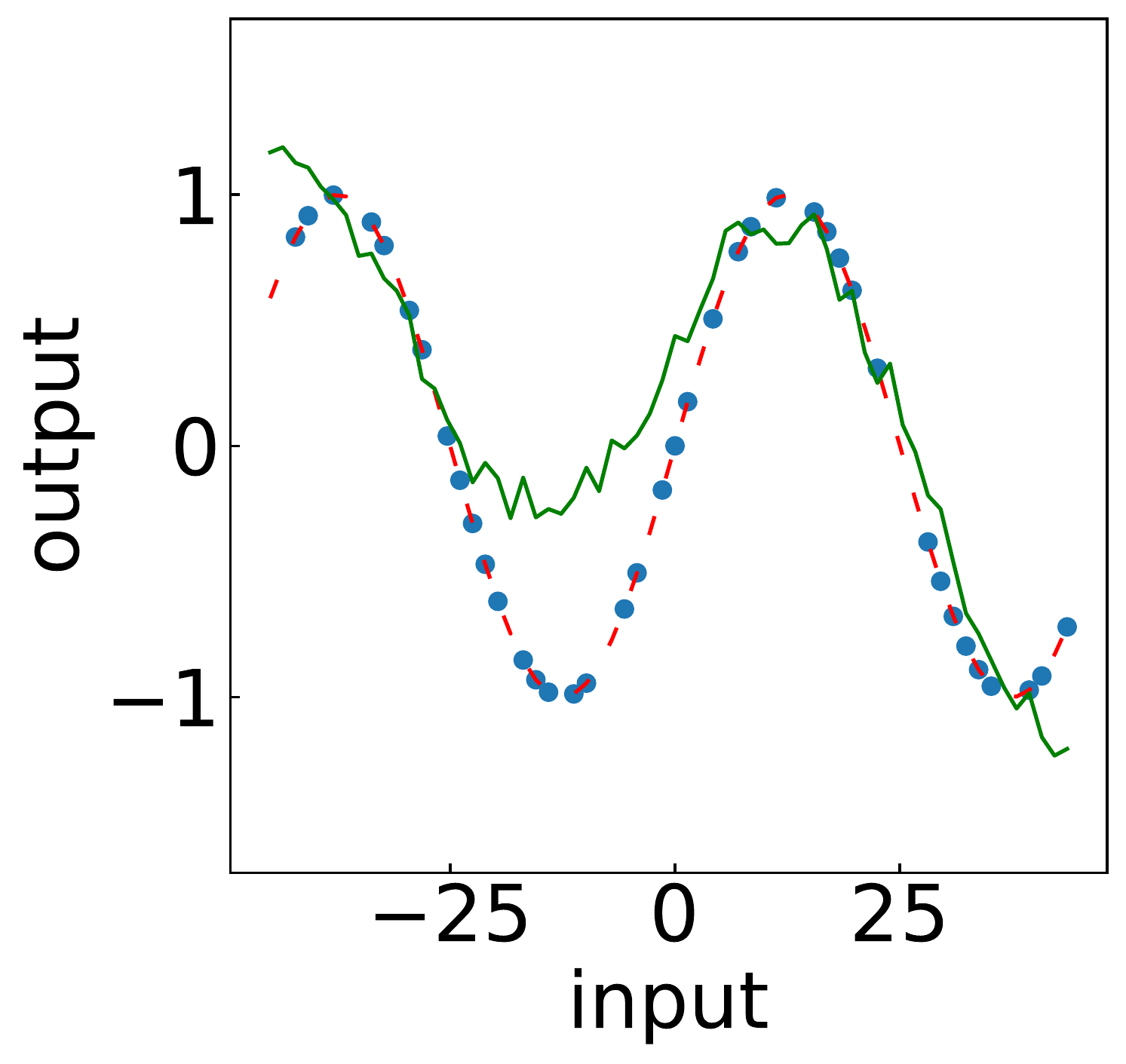}\vspace{-3truemm}
  \subcaption{N = 3}\label{sin3}
 \end{minipage}
 \begin{minipage}[b]{0.48\linewidth}
  \centering
  \includegraphics[width=37mm]
  {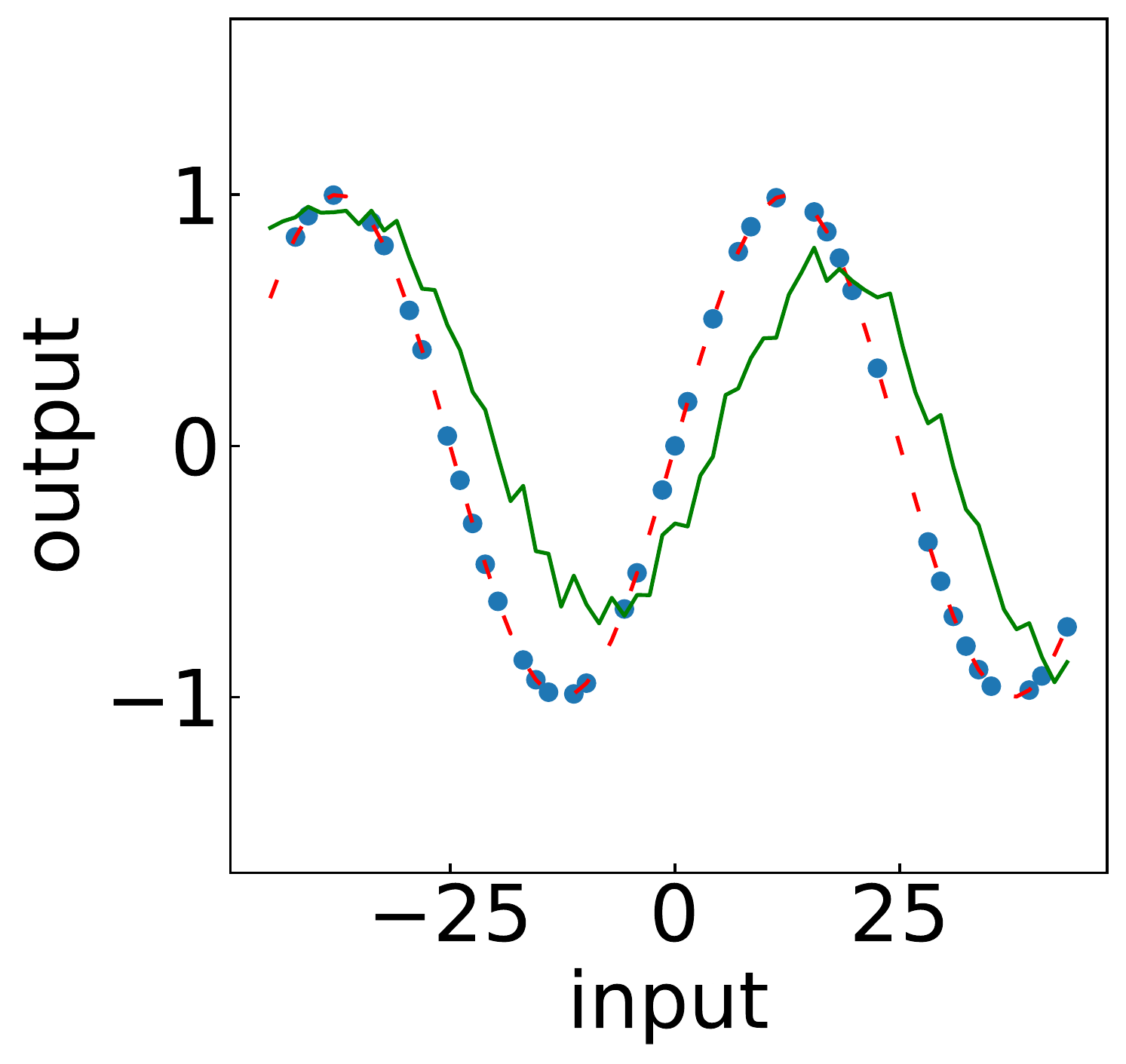}\vspace{-3truemm}
  \subcaption{N = 4}\label{sin4}
 \end{minipage}\\\vspace{1truemm}
 \begin{minipage}[b]{0.48\linewidth}
  \centering
  \includegraphics[width=37mm]
  {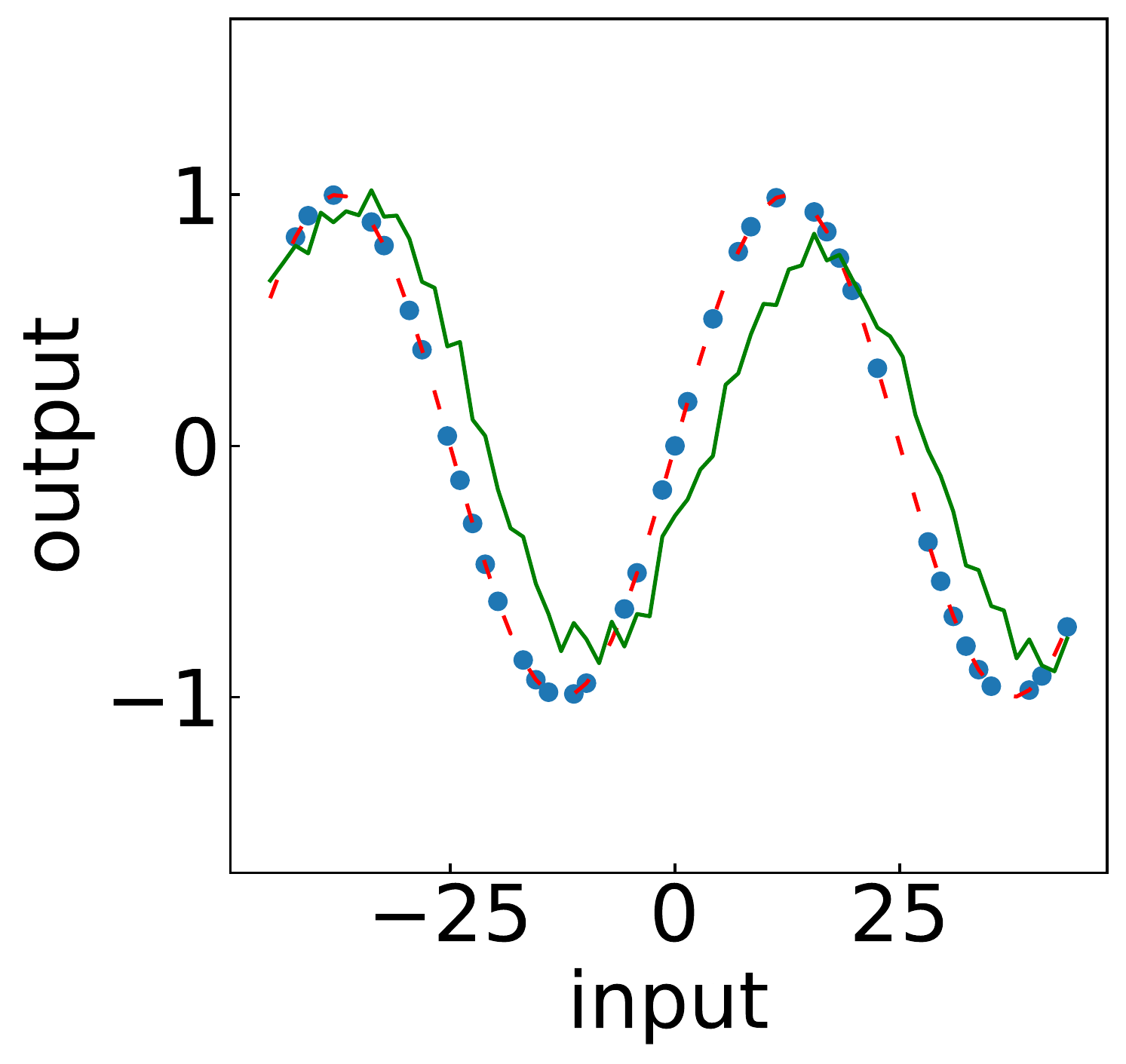}\vspace{-3truemm}
  \subcaption{N = 5}\label{sin5}
 \end{minipage}
  \begin{minipage}[b]{0.48\linewidth}
  \centering
  \includegraphics[width=37mm]
  {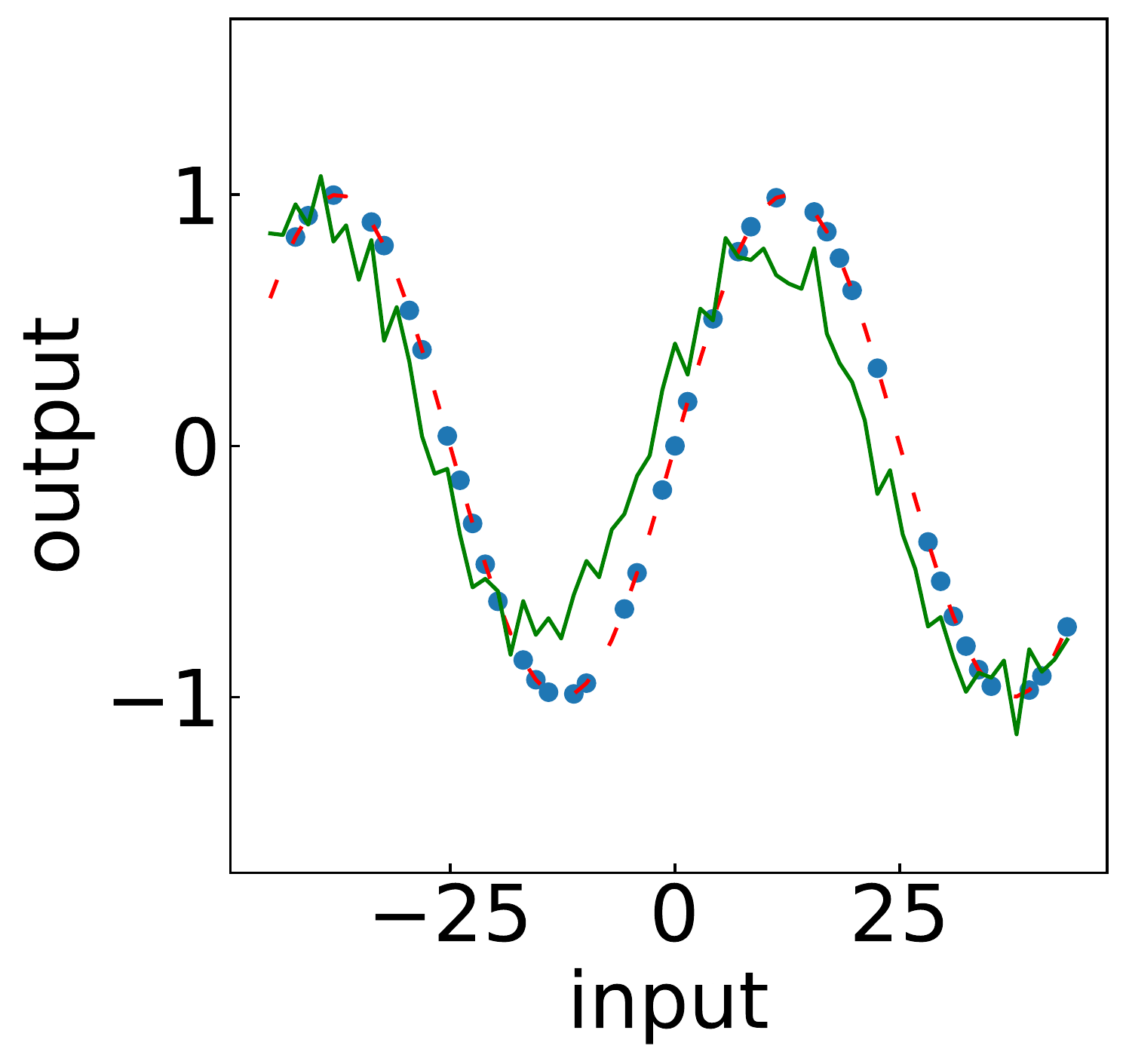}\vspace{-3truemm}
  \subcaption{N = 6}\label{sin6}
 \end{minipage}
 \end{minipage}\hspace{0.01\linewidth}
 \begin{minipage}[b]{0.49\linewidth}
 \begin{minipage}[b]{0.48\linewidth}
  \centering
  \includegraphics[width=37mm]
  {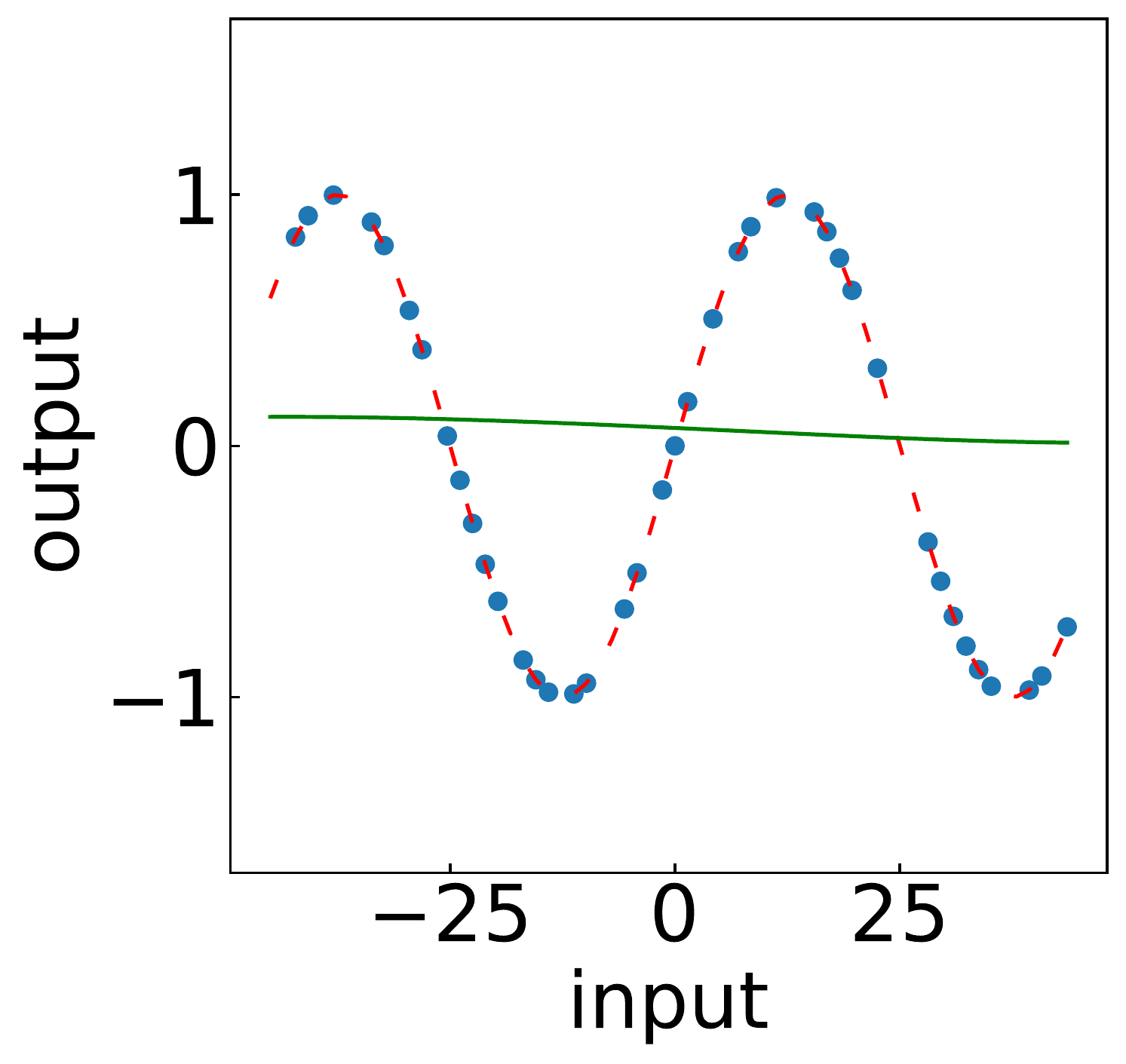}\vspace{-3truemm}
  \subcaption{$\tau=0.02$}\label{fig:sim_sin1}
 \end{minipage}
 \begin{minipage}[b]{0.48\linewidth}
  \centering
  \includegraphics[width=37mm]
  {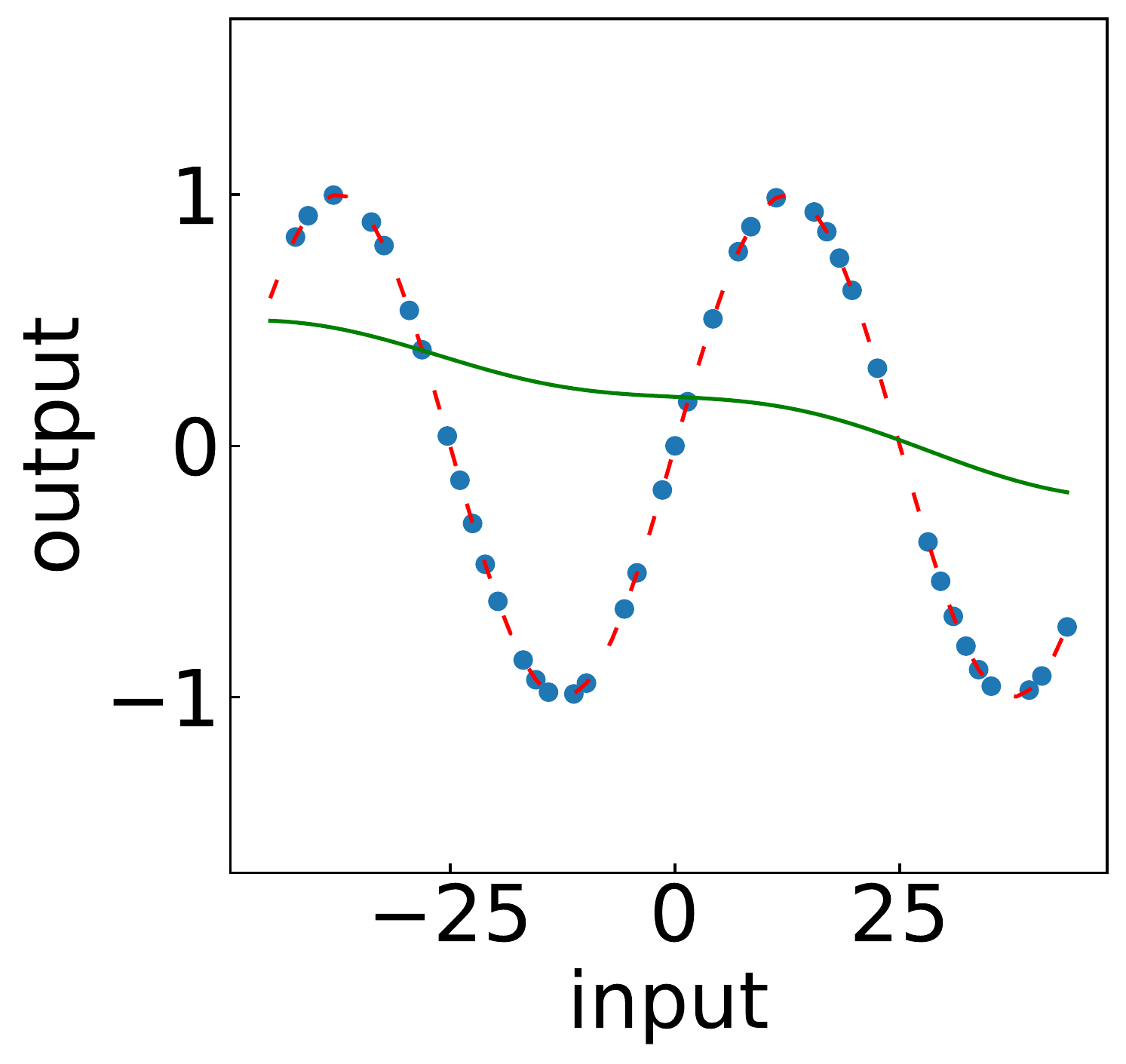}\vspace{-3truemm}
  \subcaption{$\tau=0.04$}\label{fig:sim_sin2}
 \end{minipage}\\\vspace{1truemm}
 \begin{minipage}[b]{0.48\linewidth}
  \centering
  \includegraphics[width=37mm]
  {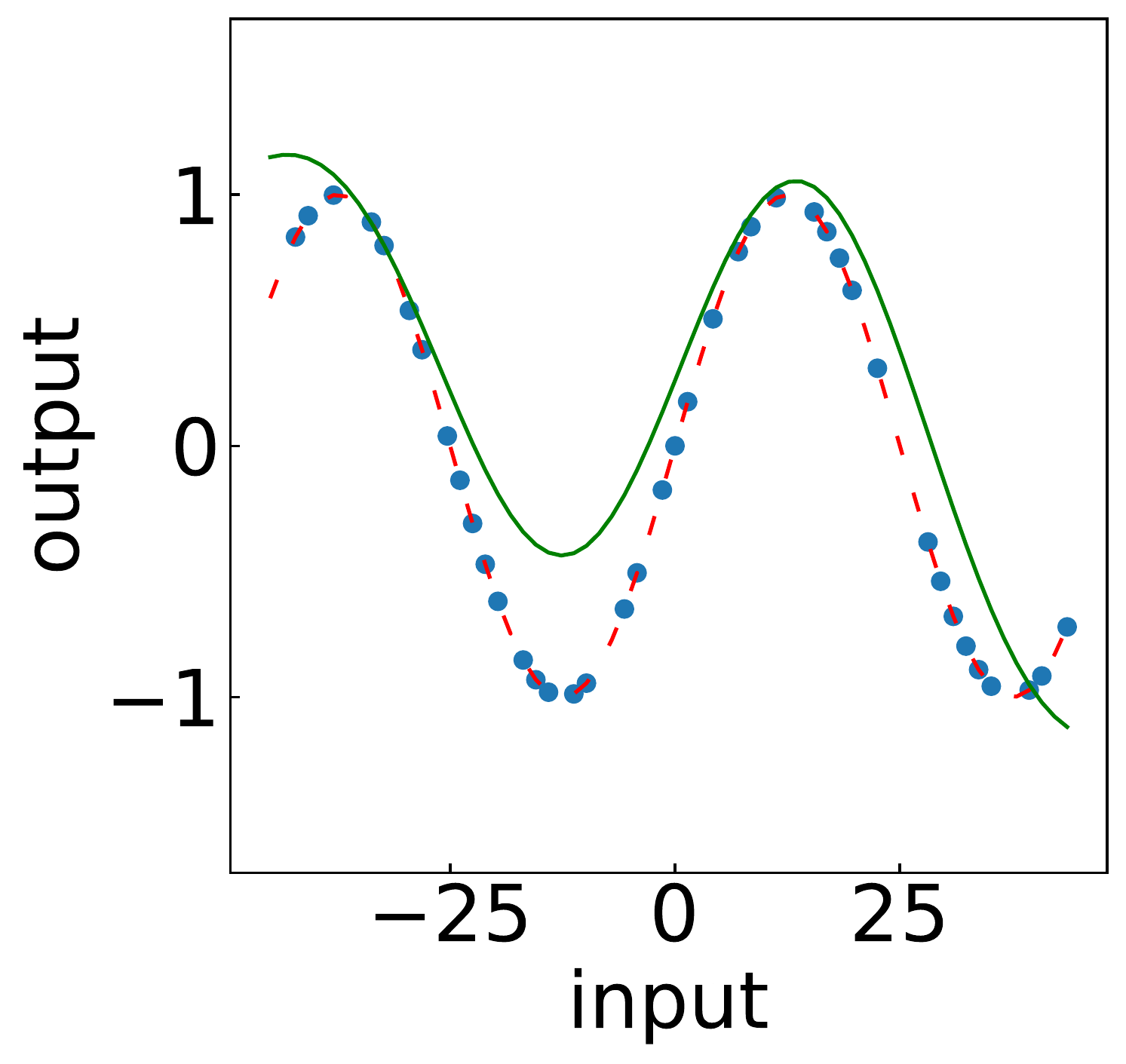}\vspace{-3truemm}
  \subcaption{$\tau=0.06$}\label{fig:sim_sin3}
 \end{minipage}
 \begin{minipage}[b]{0.48\linewidth}
  \centering
  \includegraphics[width=37mm]
  {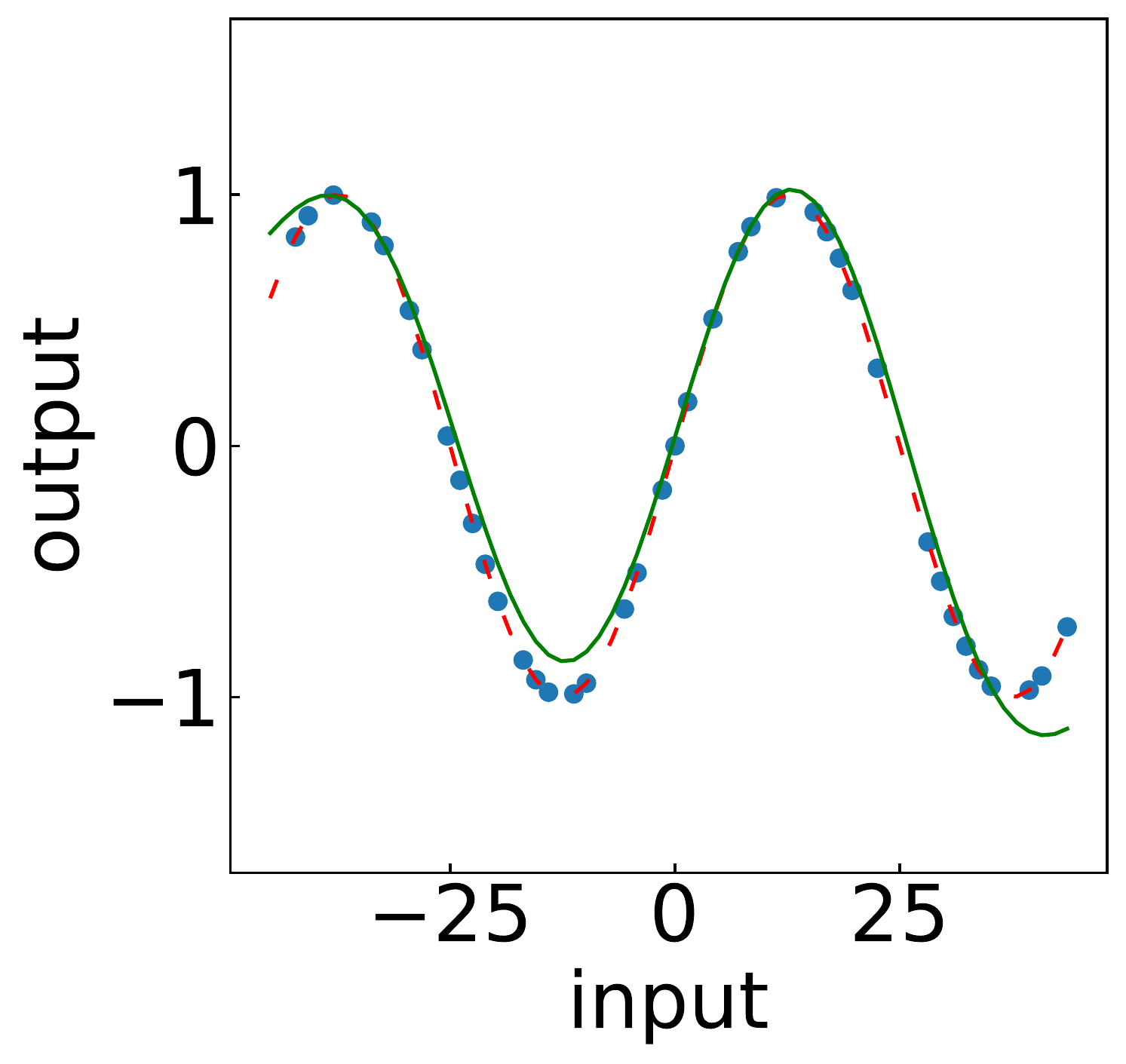}\vspace{-3truemm}
  \subcaption{$\tau=0.08$}\label{fig:sim_sin4}
 \end{minipage}\\\vspace{1truemm}
 \begin{minipage}[b]{0.48\linewidth}
  \centering
  \includegraphics[width=37mm]
  {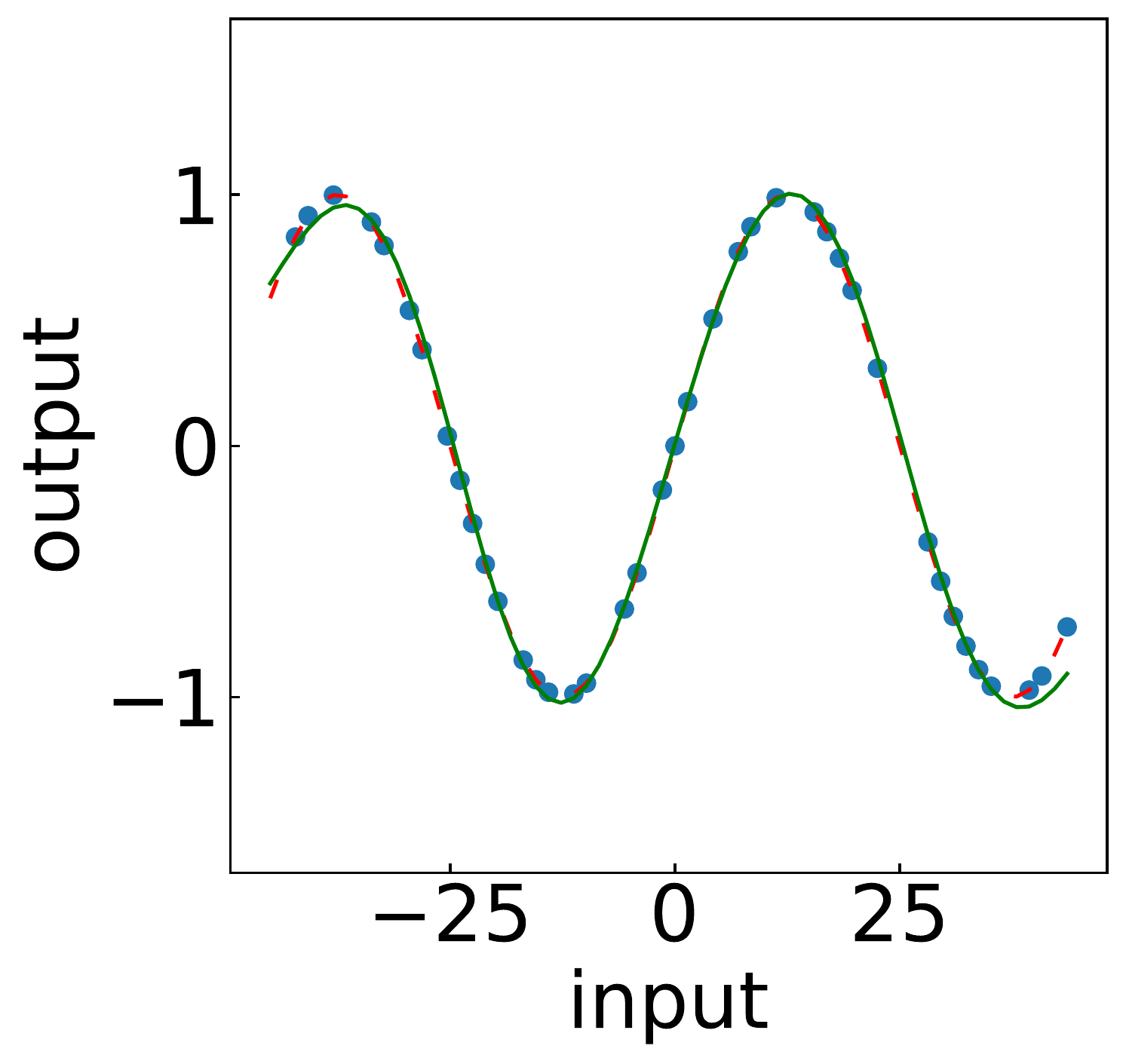}\vspace{-3truemm}
  \subcaption{$\tau=0.10$}\label{fig:sim_sin5}
 \end{minipage}
 \begin{minipage}[b]{0.48\linewidth}
  \centering
  \includegraphics[width=37mm]
  {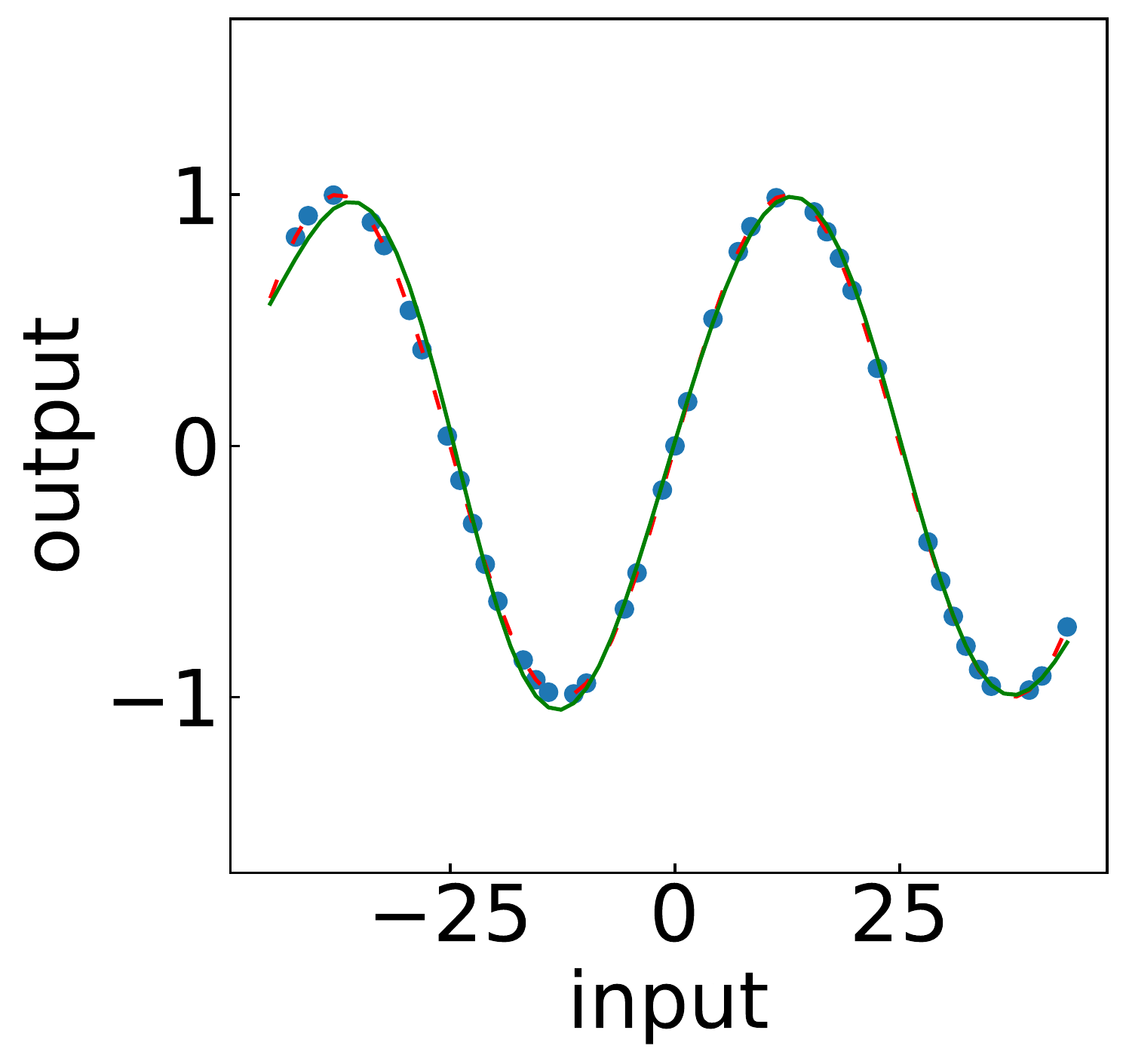}\vspace{-3truemm}
  \subcaption{$\tau=0.12$}\label{fig:sim_sin6}
 \end{minipage}
 \end{minipage}
 \caption{Demonstration of one-dimensional regression task of $y=\sin(2\pi x/50)$ performed with (a)-(f) NMR kernel and (g)-(l) numerically simulated kernel. The blue dots are the training data. The green line is the prediction of the trained model.}\label{fig:sin}
\end{figure*}

\begin{figure}
\begin{minipage}[b]{0.48\linewidth}
\centering
\includegraphics[width=\linewidth]{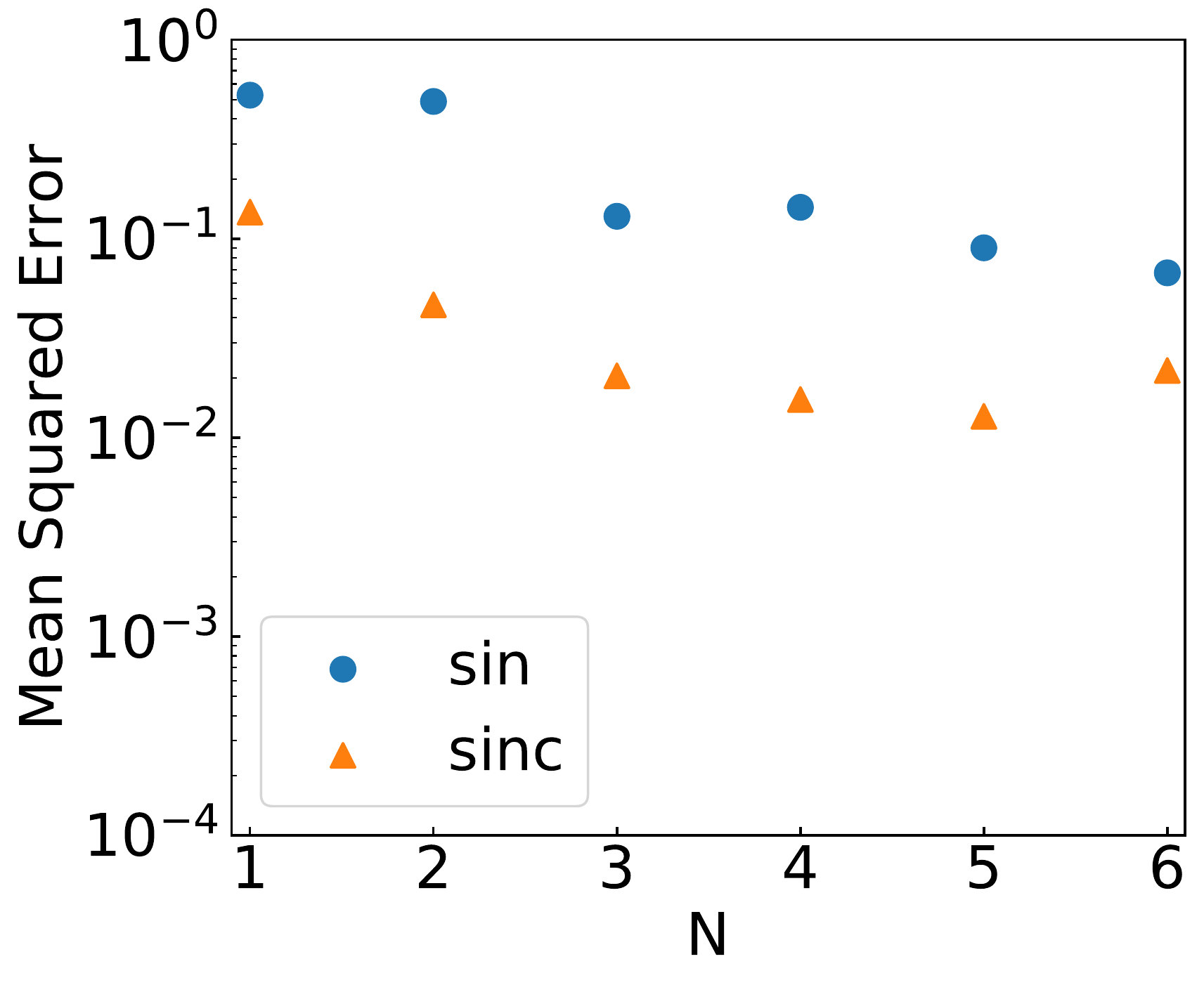}
\subcaption{}
\end{minipage}
\begin{minipage}[b]{0.48\linewidth}
\centering
\includegraphics[width=\linewidth]{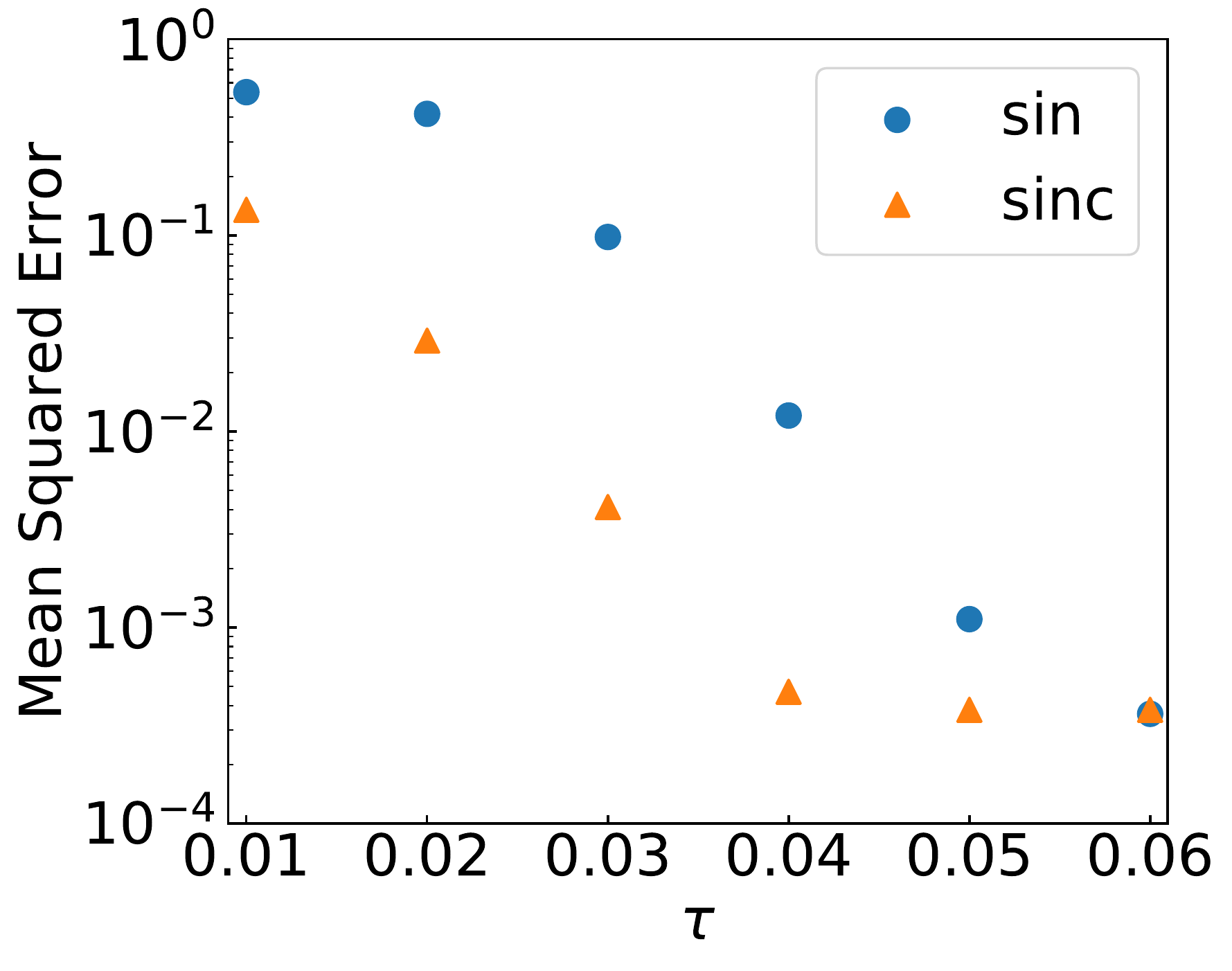}
\subcaption{}
\end{minipage}
\caption{Mean squared error of the trained models for the regression task of sin and sinc functions using (a) experimental NMR kernel and (b) numerically simulated kernel.}
\label{fig:1DMSE}
\end{figure}

\begin{figure*}[t]
\begin{minipage}[b]{0.45\linewidth}
 \begin{minipage}[b]{0.48\linewidth}
  \centering
  \includegraphics[width=37mm]
  {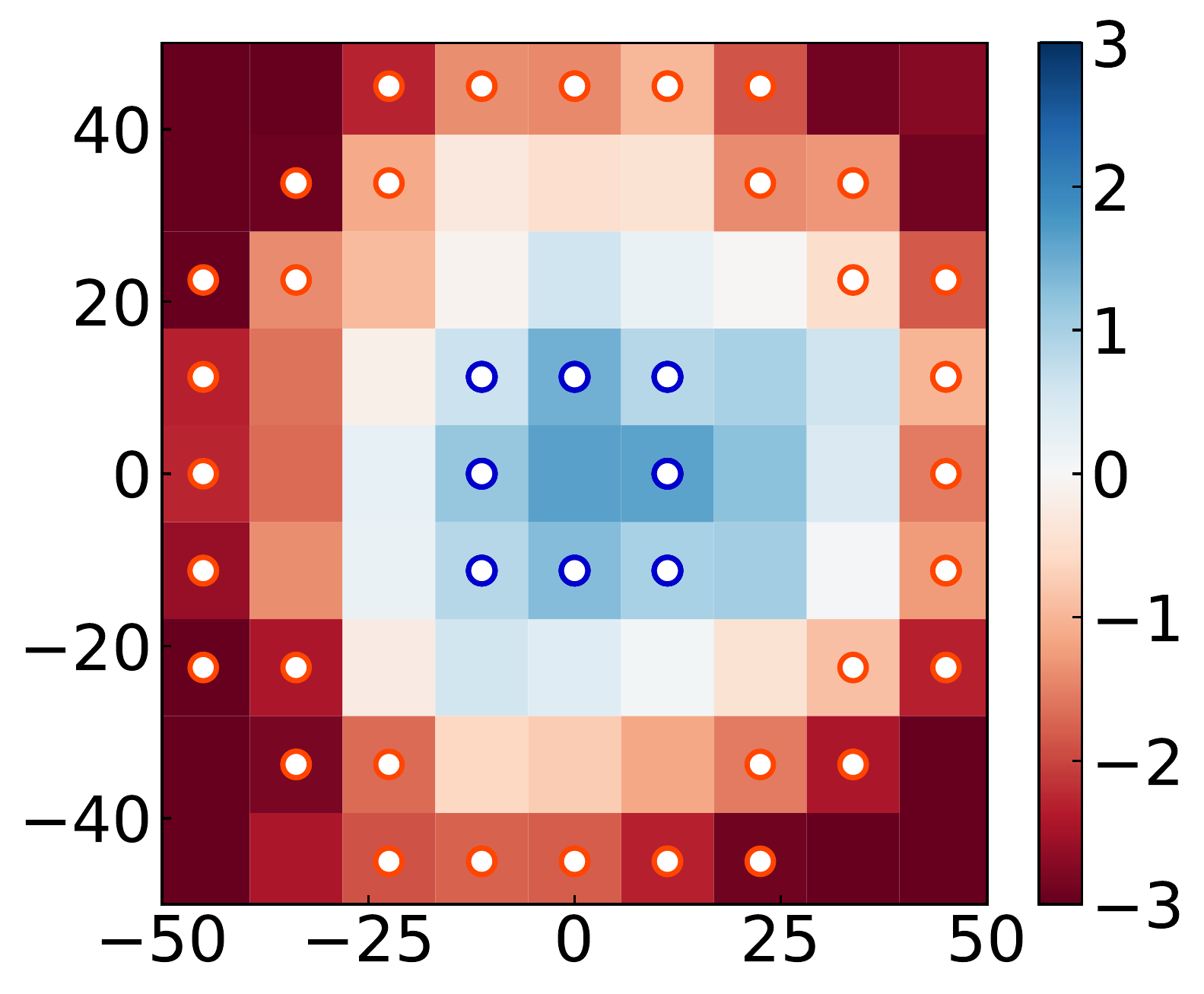}
  \vspace{-5pt}$N=1$\vspace{10pt}
 \end{minipage}
 \begin{minipage}[b]{0.48\linewidth}
  \centering
  \includegraphics[width=37mm]
  {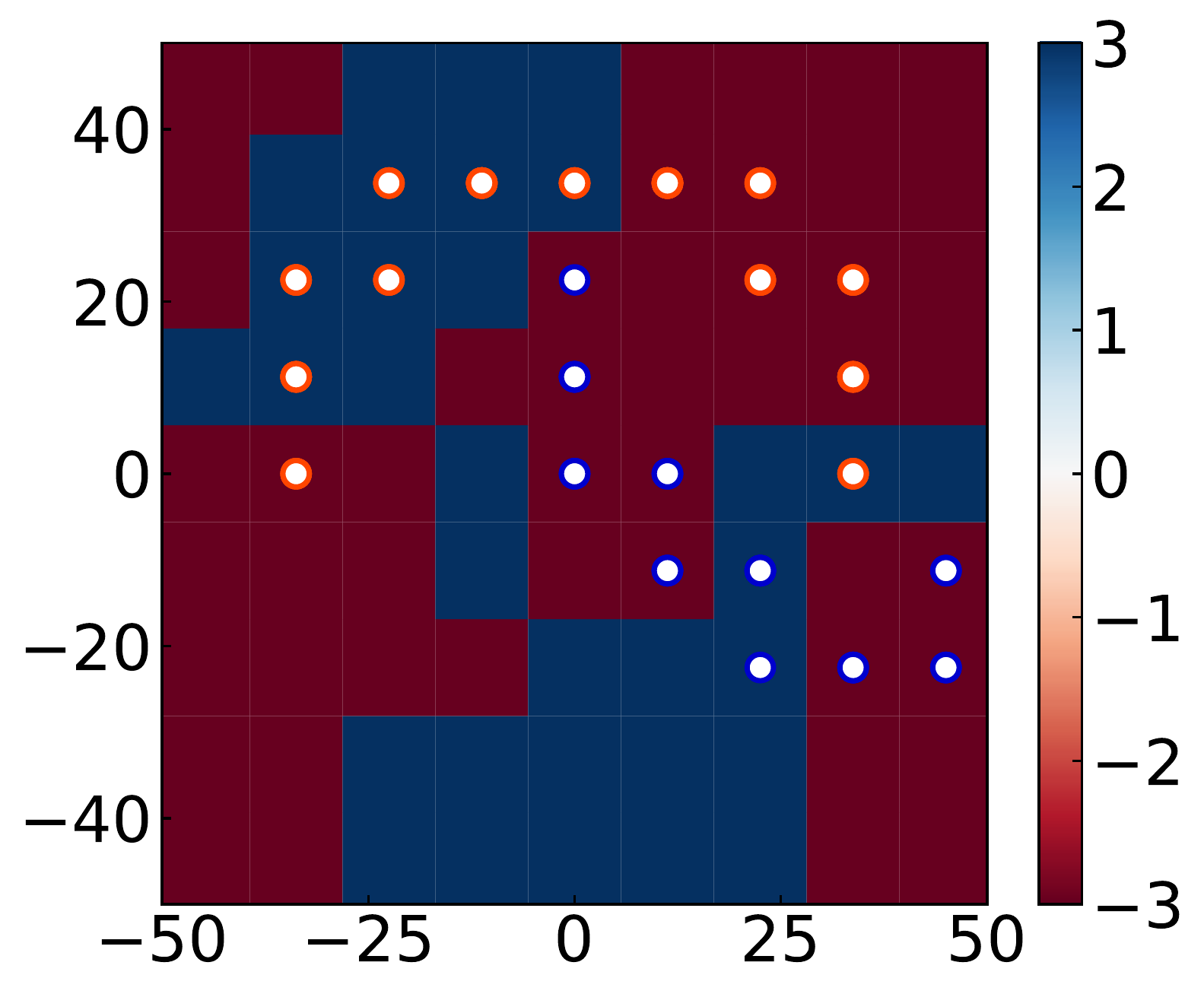}
  \vspace{-5pt}$N=1$\vspace{10pt}
 \end{minipage}
 \begin{minipage}[b]{0.48\linewidth}
  \centering
  \includegraphics[width=37mm]
  {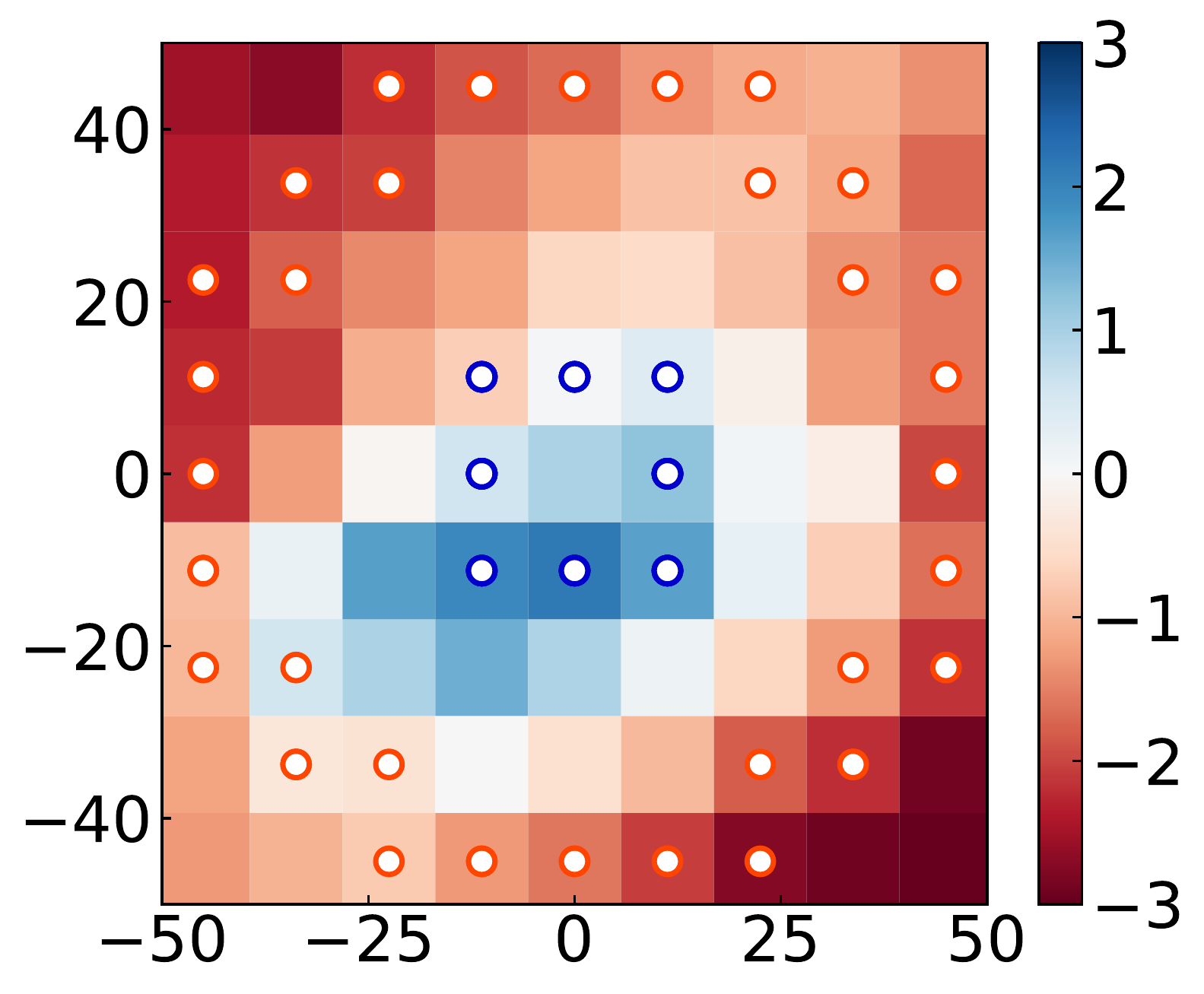}
  \vspace{-5pt}$N=2$\vspace{10pt}
 \end{minipage}
 \begin{minipage}[b]{0.48\linewidth}
  \centering
  \includegraphics[width=37mm]
  {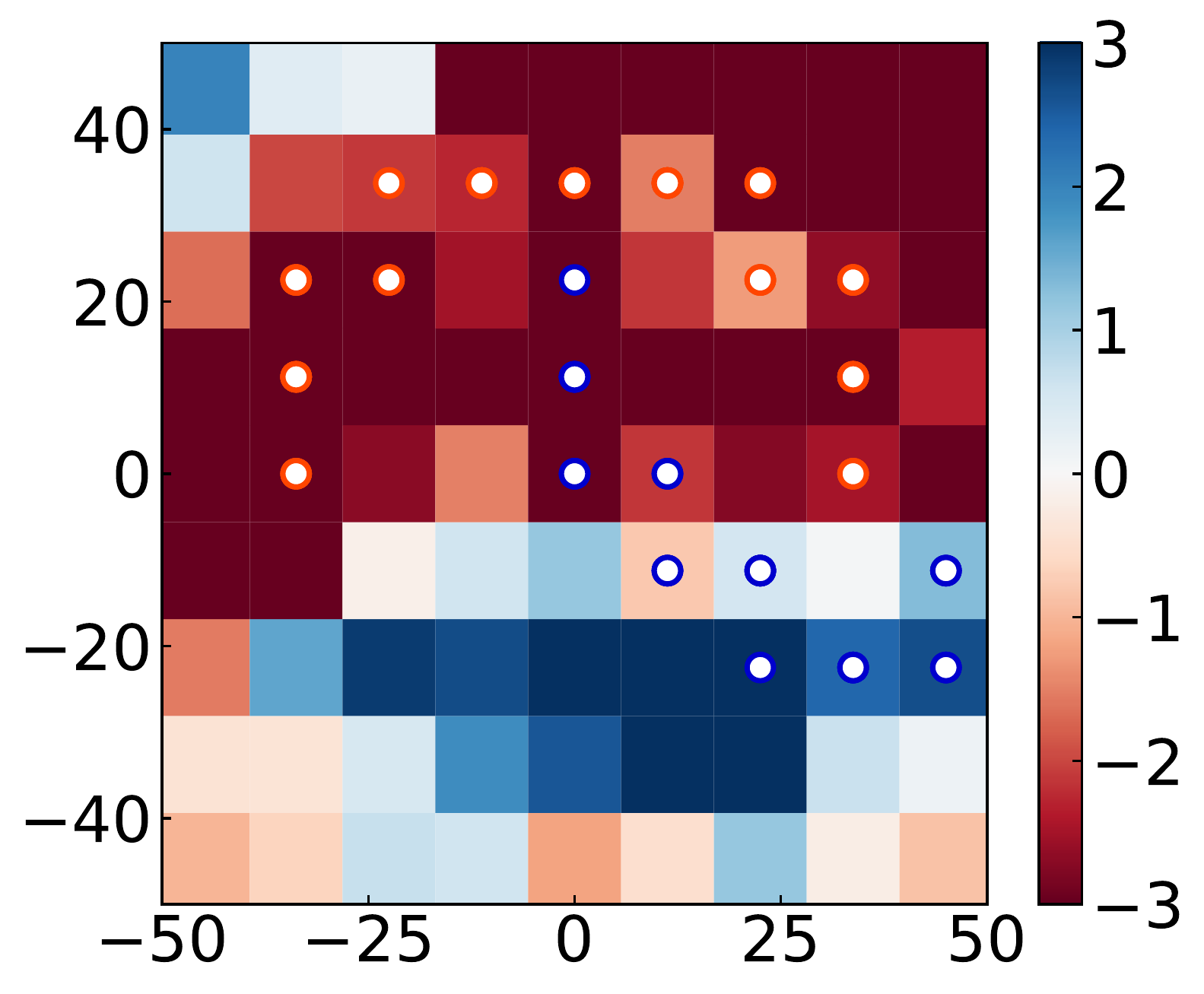}
  \vspace{-5pt}$N=2$\vspace{10pt}
 \end{minipage}
 \begin{minipage}[b]{0.48\linewidth}
  \centering
  \includegraphics[width=37mm]
  {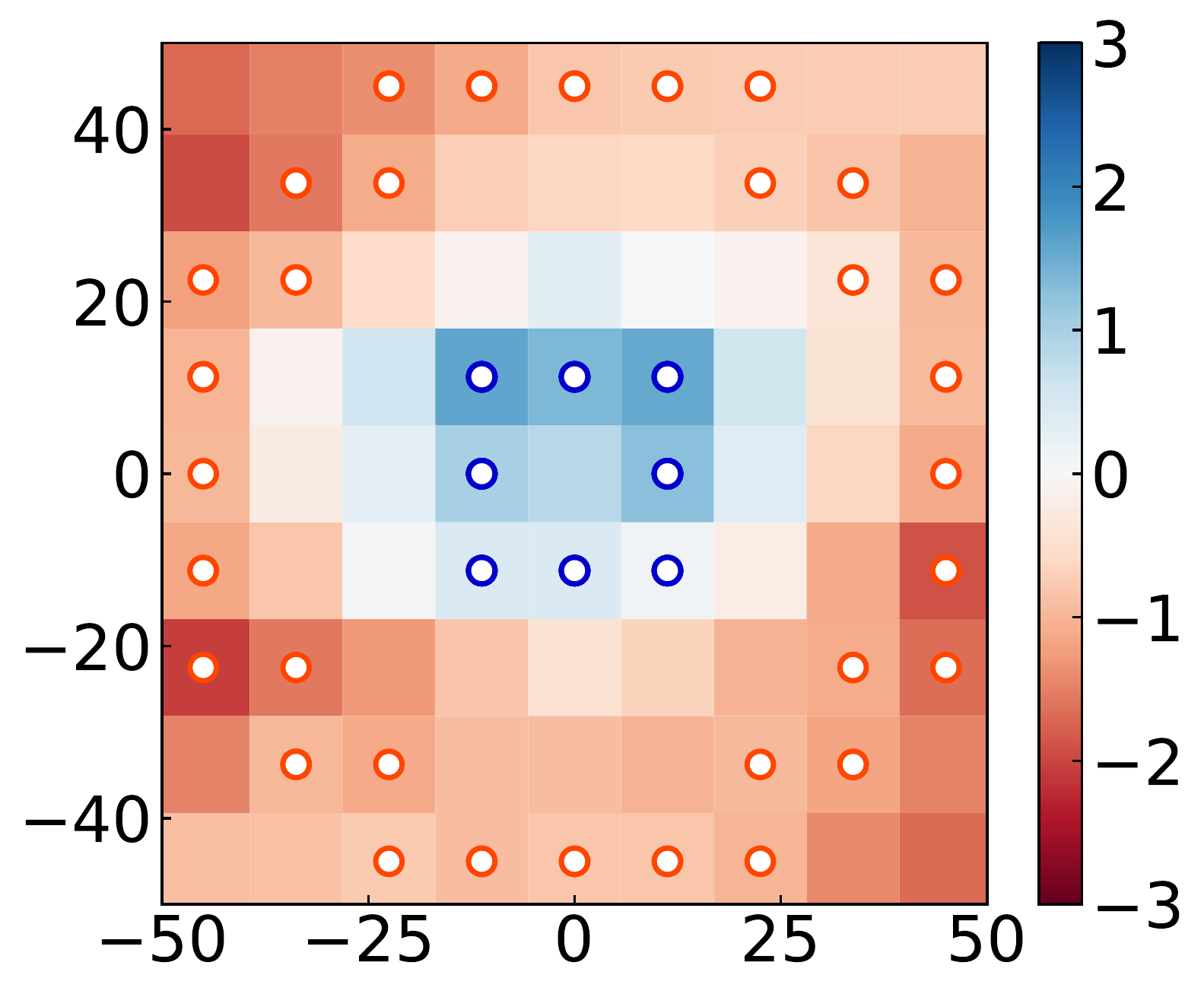}
  \vspace{-5pt}$N=3$\vspace{10pt}
 \end{minipage}
 \begin{minipage}[b]{0.48\linewidth}
  \centering
  \includegraphics[width=37mm]
  {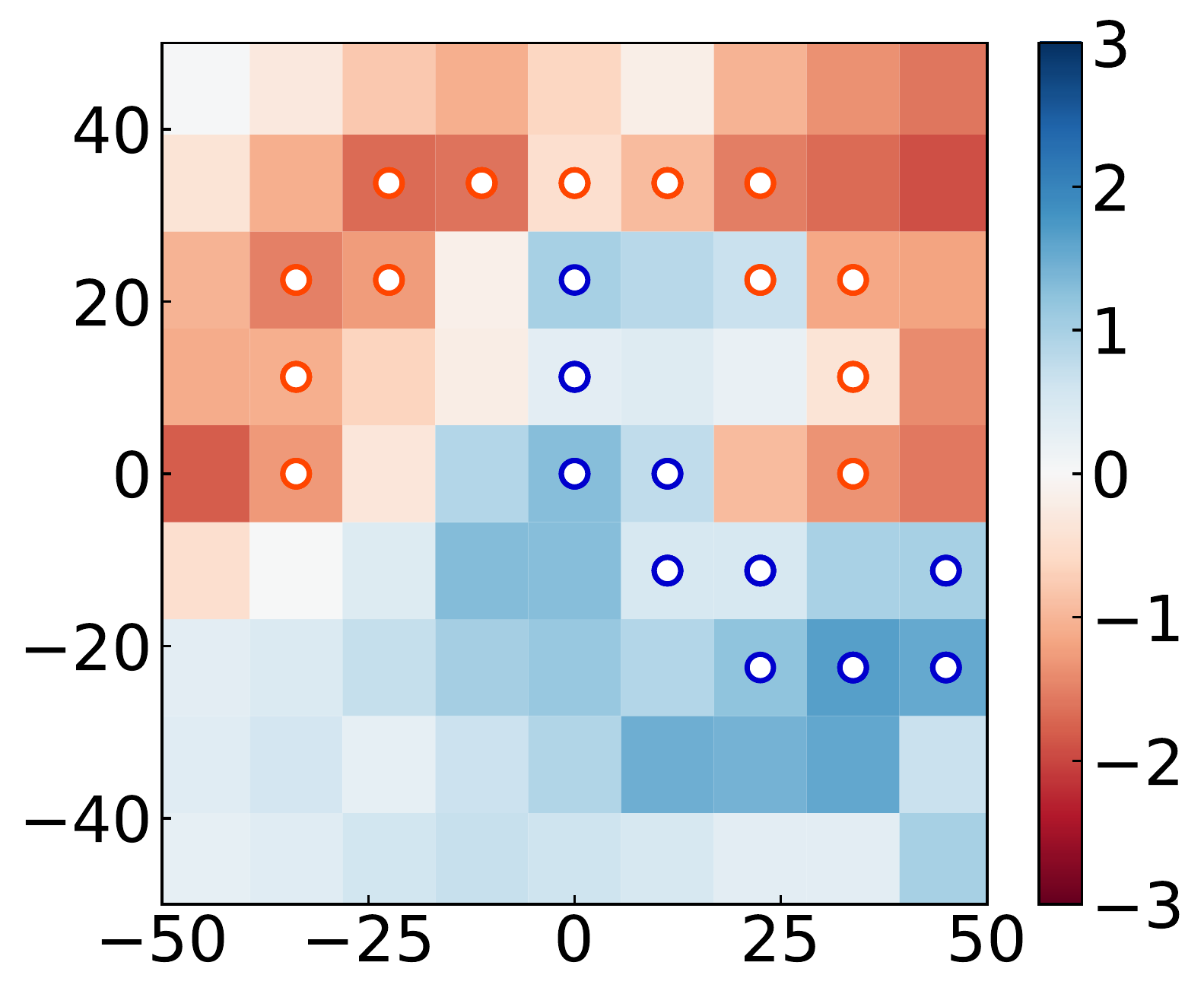}
  \vspace{-5pt}$N=3$\vspace{10pt}
 \end{minipage}
 \subcaption{}
 \end{minipage}\hspace{0.08\linewidth}
 \begin{minipage}[b]{0.45\linewidth}
 \begin{minipage}[b]{0.48\linewidth}
  \centering
  \includegraphics[width=37mm]
  {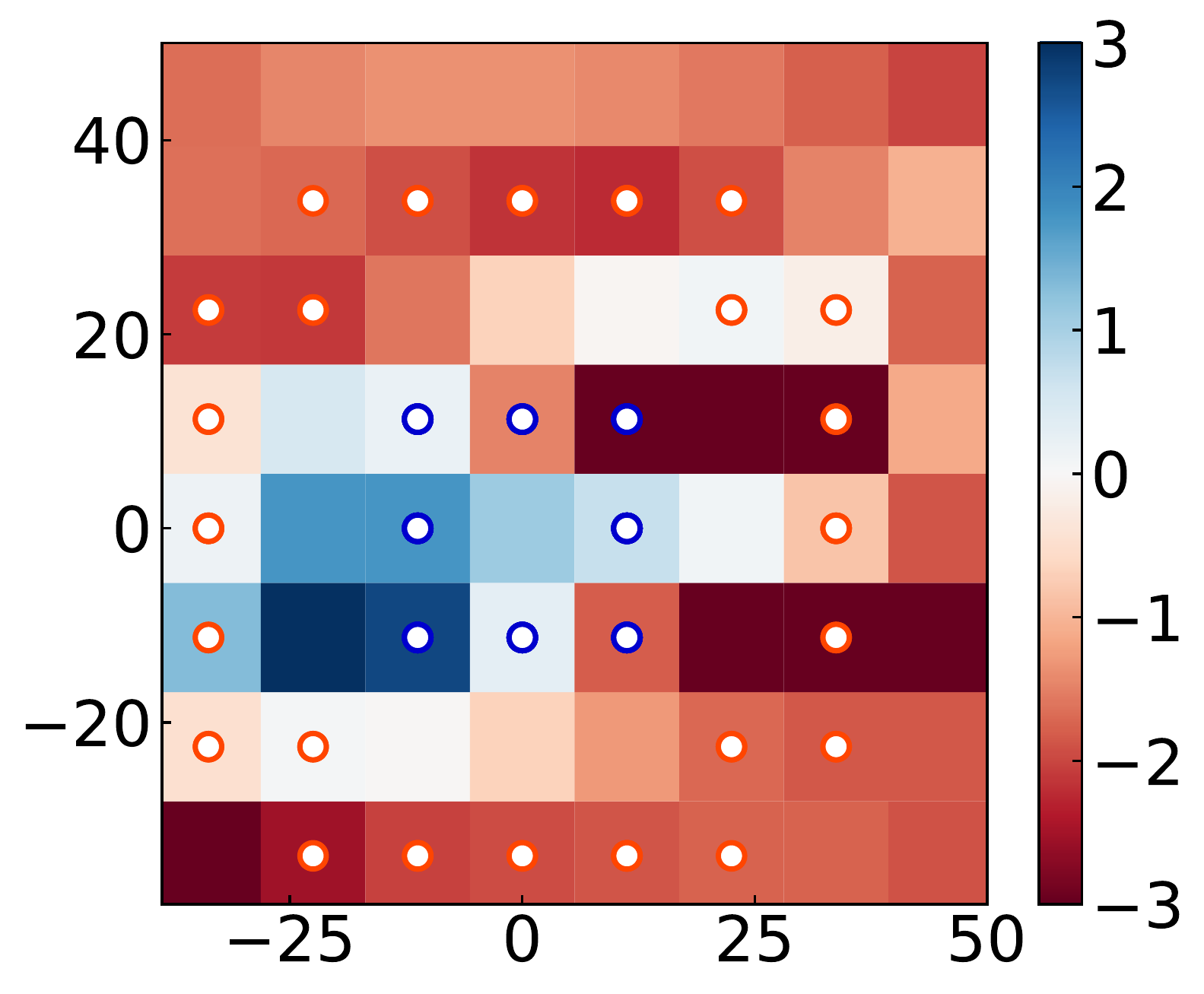}
  \vspace{-5pt}$\tau=0.03$\vspace{10pt}
 \end{minipage}
 \begin{minipage}[b]{0.48\linewidth}
  \centering
  \includegraphics[width=37mm]
  {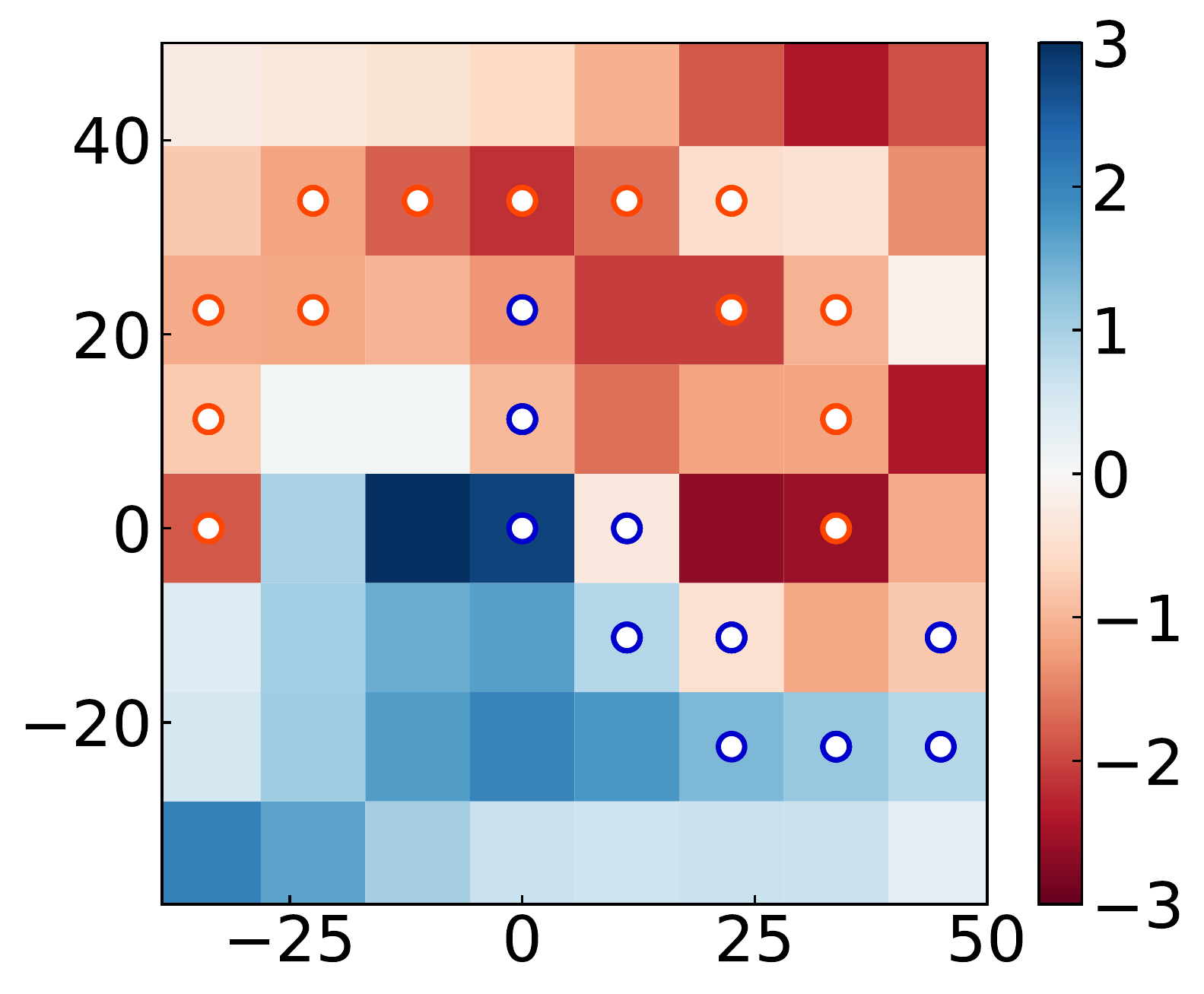}
  \vspace{-5pt}$\tau=0.03$\vspace{10pt}
 \end{minipage}
 \begin{minipage}[b]{0.48\linewidth}
  \centering
  \includegraphics[width=37mm]
  {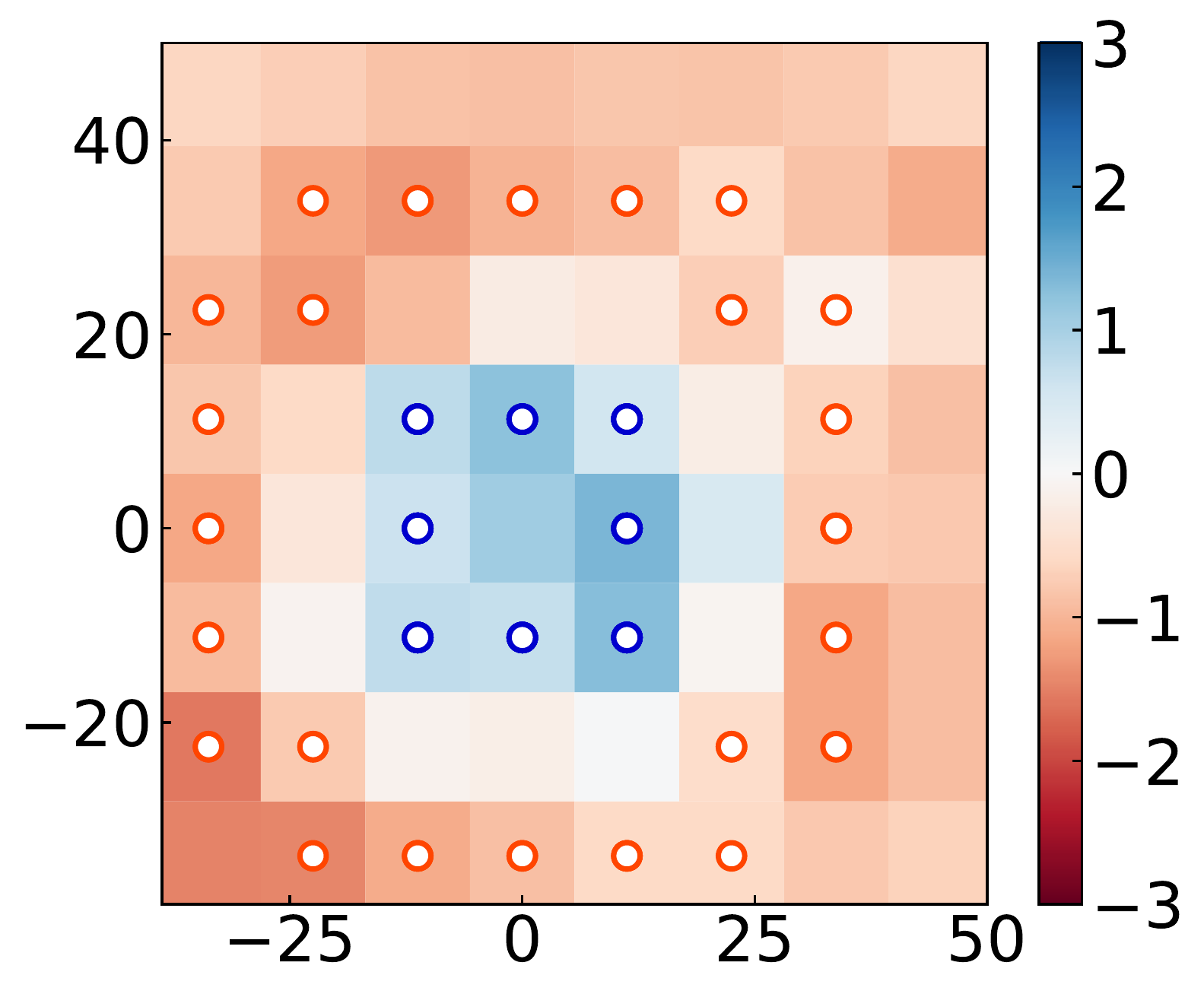}
  \vspace{-5pt}$\tau=0.06$\vspace{10pt}
 \end{minipage}
 \begin{minipage}[b]{0.48\linewidth}
  \centering
  \includegraphics[width=37mm]
  {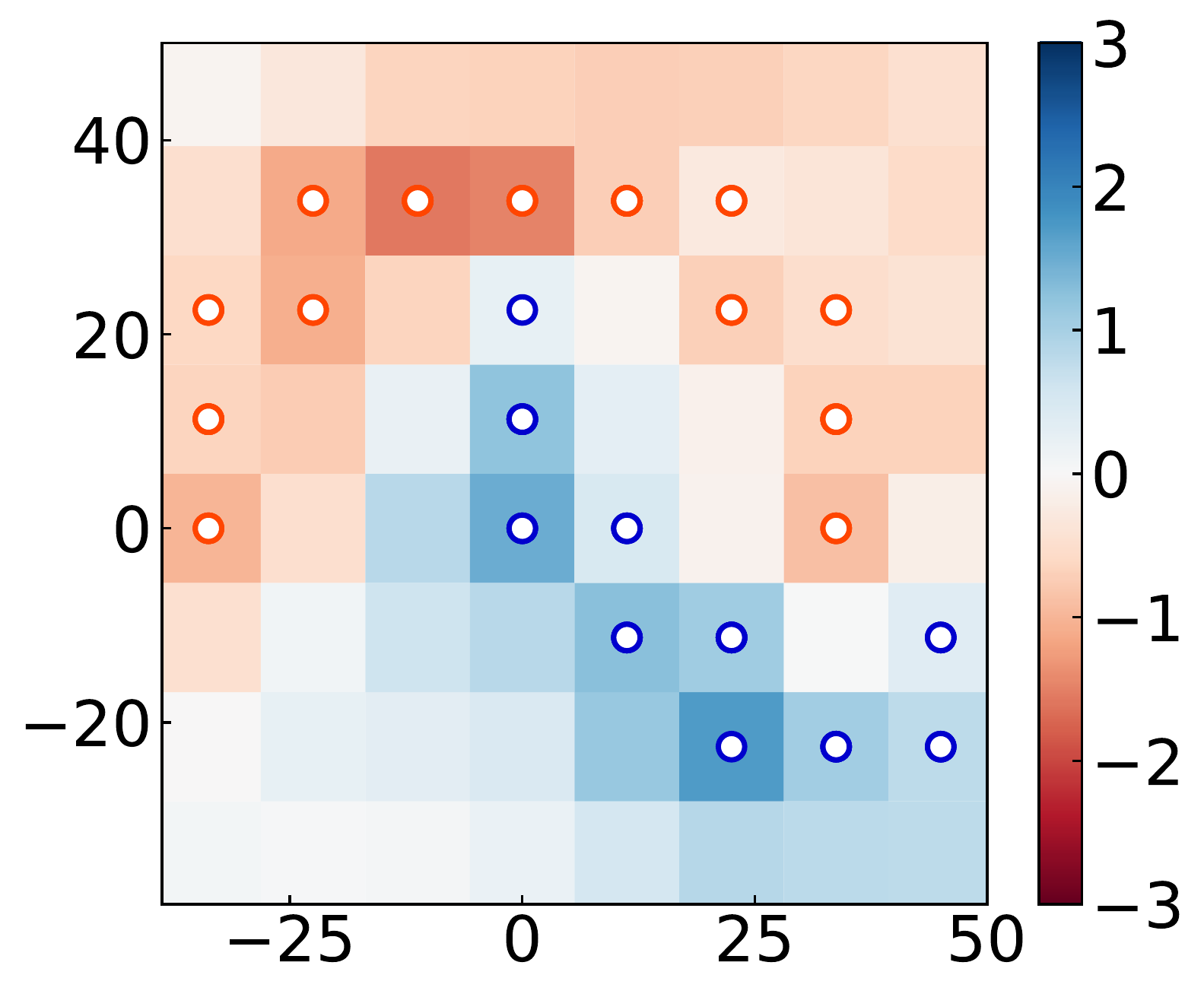}
  \vspace{-5pt}$\tau=0.06$\vspace{10pt}
 \end{minipage}
 \begin{minipage}[b]{0.48\linewidth}
  \centering
  \includegraphics[width=37mm]
  {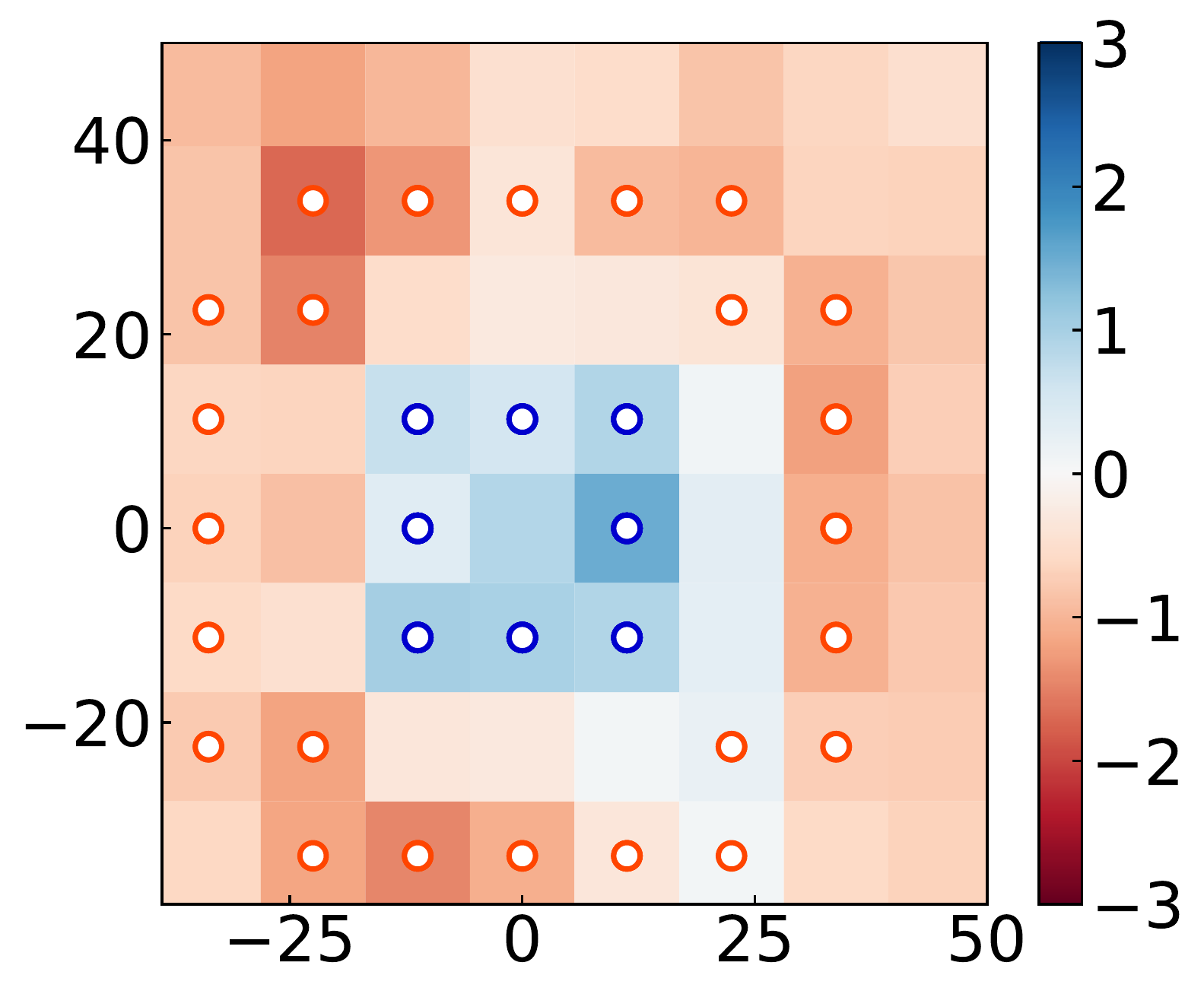}
  \vspace{-5pt}$\tau=0.09$\vspace{10pt}
 \end{minipage}
 \begin{minipage}[b]{0.48\linewidth}
  \centering
  \includegraphics[width=37mm]
  {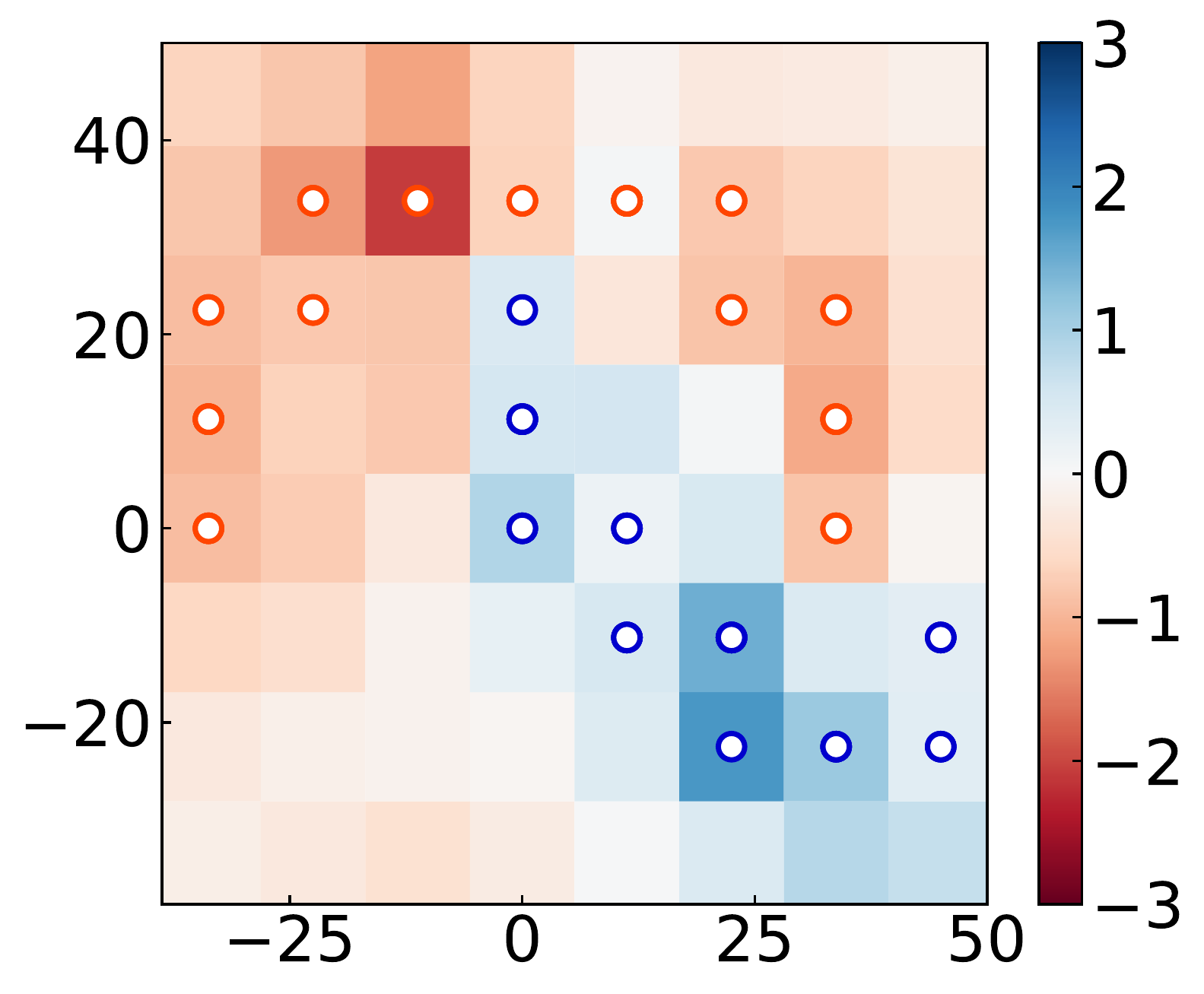}
  \vspace{-5pt}$\tau=0.09$\vspace{10pt}
 \end{minipage}
 \subcaption{}
 \end{minipage}
 \caption{(a) Classification with the NMR kernel.
 Left and right panels respectively show the results for ``circle'' and ``moon'' dataset. The dots in the figures represent training data.
 The background color indicates the decision function of the trained model. 
 Top, middle, and bottom panels are the results with $N = 1, 2, 3$ NMR kernels, respectively.
 (b) Results from the numerically simulated quantum kernel. Top, middle, and bottom panels are the results with simulated kernel with $\tau = 0.03, 0.06, 0.09$. All the other notations follow that of (a).}\label{fig:2D_classification}
\end{figure*}

\subsubsection*{Numerical simulation}
To certify the trend without the effect of noise, we conducted numerical simulations of $20$-qubit dynamics.
We drew the interaction strength, $d_{\mu\nu}$, from uniform distribution on $[-1,1]$ for all $\mu,\nu$.
The evolution according to the Hamiltonian $H(x)$ is approximated by the first-order Trotter formula, that is,
\begin{align}
    e^{-iH(x)\tau} \approx e^{-ix I_z} \left[\prod_{\mu<\nu}  e^{-i\tau d_{\mu,\nu}(I_{y,\mu}I_{y,\nu}-I_{x,\mu}I_{x,\nu})/M}\right]^M e^{ix I_z}.
\end{align}
We set $\tau=0.01,0.02,\cdots,0.06$ and $\tau/M=0.001$ in the simulation.
In order to reduce the computational cost, we set $A(\bm{x}) = U(\bm{x})\prod_{\mu}\left(\frac{I}{2}+I_{z,\mu}\right)U^\dagger (\bm{x})$, which allows us to evaluate $\Tr(A(\bm{x}_i)A(\bm{x}_j))$ by computing $\left|\bra{0}U^\dagger(\bm{x}_j)U(\bm{x}_i)\ket{0}\right|^2$, where $\ket{0}$ is the ground state of $I_z$.
Since we can compute this quantity by simulating dynamics of a $2^{20}$-dimensional state vector, it is significantly easier than computing $A(\bm{x}) = U(\bm{x}) I_z U^\dagger (\bm{x})$ where we would need to simulate dynamics of $2^{20}\times 2^{20}$ matrices.
All simulations are performed with a quantum circuit simulator Qulacs \cite{Qulacs}.
The result for the sin function is shown as Figs.~\ref{fig:sin} (g)-(l).
For the sinc function, we place the result in Supplementary Material.
The mean squared error of the prediction evaluated in the same manner is shown in Fig. \ref{fig:1DMSE} (b).
We can see the performance gets better With increasing $\tau$, which corresponds to increasing $N$ in the experiment.
This certifies the trend observed in the NMR experiment.

\subsection{Two-dimensional classification task}
As the second demonstration, we implement two-dimensional classification tasks.
We employ the hard-margin kernel support vector machine \cite{Bishop2011} and its implementation in scikit-learn \cite{scikit-learn} for this task.
The training data set is generated by the circle and moon dataset of scikit-learn \cite{scikit-learn}.
We used the NMR kernel with $N=1, 2, 3$.
We again conducted numerical simulations with the same setting as the previous section along with the experiment with $\tau=0.03, 0.06, 0.09$.
The value of this numerical kernel is shown in the Supplementary Material.

The results are shown in Fig.~\ref{fig:2D_classification}.
We note that, for the moon dataset with $N=1$ experimental NMR kernel, the kernel matrix was singular, and we did not obtain a reliable result.
We reason this to the broadness of the kernel at $N=1$.
For these classification tasks, we do not observe particular changes with increased evolution time.
It is also clear from the hinge loss, which is a measure of the accuracy in classification tasks, for each result given in Supplementary Material.

\section{Discussion}
In the one-dimensional regression task, we observed the trend of better performance with longer evolution time.
This can be explained by the shape of the kernel generated by the NMR dynamics, which is shown in Fig. \ref{fig:1DKernel} (c).
As mentioned earlier, this experiment is essentially the Loschmidt echo, and the shape of the signal sharpens as the evolution time increases.
The sharpness of the kernel can directly be translated to the representability of the model as it can be in the popular gaussian kernel, because this property allows the machine to distinguish different data more clearly.
However, it also causes overfitting problems if the data points are sparse.
The most extreme case is when we use delta function as a kernel, where every training point is learned with the perfect accuracy while the trained model fails to predict for unknown inputs.
In our experimental case, we did not observe any overfitting problem, which means that our training samples were dense enough for the sharpness of the kernel utilized in the model, and thus we observed an increasing performance from the improved representability of the kernel with longer evolution time.
On the other hand, for the classification task, we did not observe any significant trend depending on the evolution time.
We suspect that there is an optimal evolution time for this kind of task, which should be explored in future works.

We note that the shape of the kernel resembles the gaussian kernel which is widely employed in many machine learning tasks.
One might think that we could have obtained similar results using the gaussian kernel.
While this is true, it is also true that the NMR dynamics evolved by general Hamiltonian cannot be simulated classically under certain complexity conjecture \cite{Morimae2014, Fujii2018}.
This leaves a possibility that the NMR kernel performs better than classical kernel in some specific cases.
More experiments using different Hamiltonians are required to test whether the ``quantum'' kernel has any advantage in machine learning tasks over widely used conventional kernels.

\section{Conclusion}
We proposed and experimentally tested a quantum kernel expressed by  Eqs.~(\ref{eq:qkernel}) and (\ref{eq:uforkernel}).
Experimentally, we used $^1\mathrm{H}$ spins in adamantane with $O(10)$ coherence order to compute the kernel.
Machine learning models for one-dimensional regression tasks and two-dimensional classification tasks were constructed with the proposed kernel.
The experimental and numerical results showed similar results. 
Experiments along with numerical simulation also showed that the performance of the model tended to increase with longer evolution time, or equivalently, with a larger number of spins involved in the dynamics for certain tasks.
It would be interesting to export this method to more quantum-oriented machine learning tasks.
For example, one may be able to distinguish two dynamical phases of spin systems, such as localized and delocalized phases demonstrated in Ref.~\cite{Alvarez2015}, with the kernel support vector machine employed in this work.
More experiments are needed to verify the power of this ``quantum kernel'' approach, but our results can be thought of as one of the baselines of this emerging field.

\begin{acknowledgements}
    KM thanks the METI and IPA for their support through the MITOU Target program. KM is also supported by JSPS KAKENHI No. 19J10978. KF is supported by KAKENHI No.16H02211, JST PRESTO JPMJPR1668, JST ERATO JPMJER1601, and JST CREST JPMJCR1673. MN is supported by JST PRESTO JPMJPR1666. This work is supported by MEXT Quantum Leap Flagship Program (MEXT Q-LEAP) Grant Number JPMXS0118067394.
\end{acknowledgements}

\clearpage
\begin{center}
\textbf{\large Supplementary Material}
\end{center}
\setcounter{equation}{0}
\setcounter{figure}{0}
\setcounter{table}{0}
\setcounter{section}{0}
\setcounter{page}{1}
\makeatletter
\renewcommand{\theequation}{S\arabic{equation}}
\renewcommand{\thefigure}{S\arabic{figure}}

\section{Experimental details}
\begin{figure}[H]
    \centering
    \includegraphics[width=0.6\linewidth]{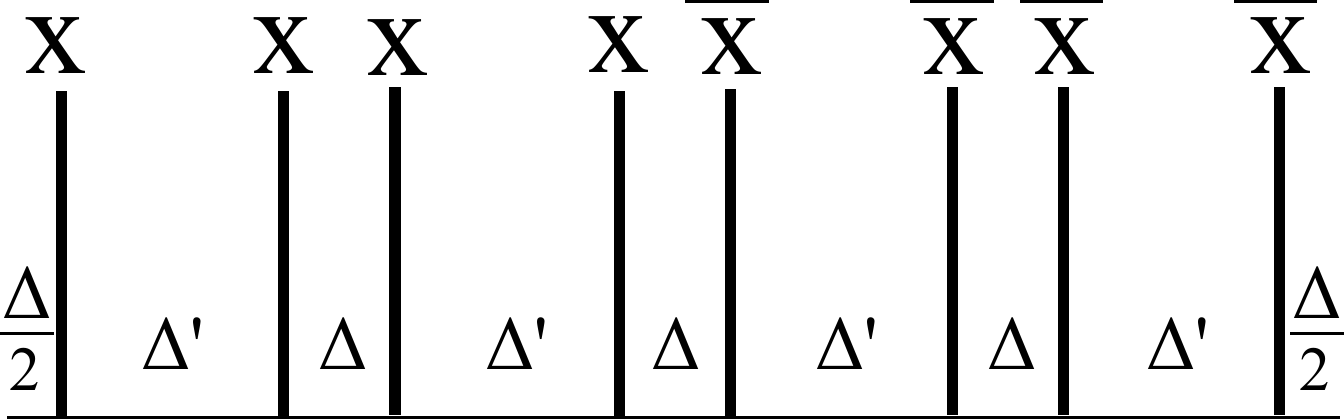}
    \caption{Pulse sequence to generate $H(0)$. $X$ and $\overline{X}$ respectively stand for $\pi/2$ and $-\pi/2$ rotation around $x$-axis.}
    \label{fig:8pulse}
\end{figure}

A pulse sequence to realize $H(0)$ is given in Fig.~\ref{fig:8pulse}.
In the experiment, we set the length of $\pi/2$ pulse, $\tau_p$, to $1.5~\mu s$. For the waiting period, we used $\Delta'=2\Delta+\tau_p$ with $\Delta = 3.5~\mu s$, which makes the evolution time for a cycle, $\tau_1$, $60~\mu s$.
By repeating the sequence for $N$ times, we can effectively evolve the spins with $e^{-iH(x_D)\tau}$ for $\tau=N\tau_1$.
NMR spectroscopy with polycrystalline adamantane sample was performed at room temperature with OPENCORE NMR \cite{TAKEDA2008}, operating at a resonant frequency of 400.281 MHz for $^1\text{H}$ nucleus observation.

\section{One-dimensional Regression task}

\begin{figure*}
 \begin{minipage}[b]{0.48\linewidth}
 \begin{minipage}[b]{0.48\linewidth}
  \centering
  \includegraphics[width=37mm]
  {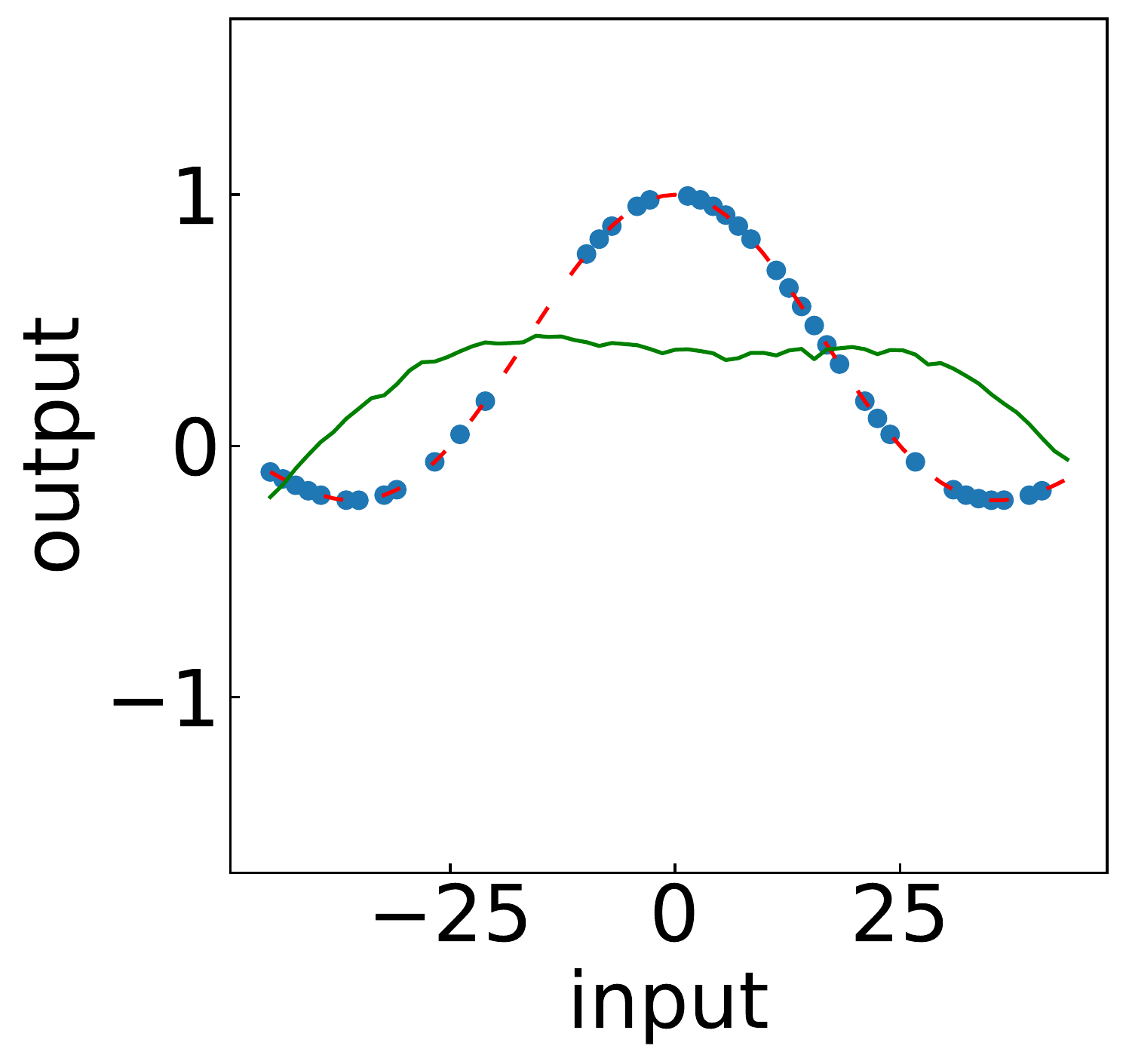}\vspace{-3truemm}
  \subcaption{N = 1}\label{sinc1}
 \end{minipage}
 \begin{minipage}[b]{0.48\linewidth}
  \centering
  \includegraphics[width=37mm]
  {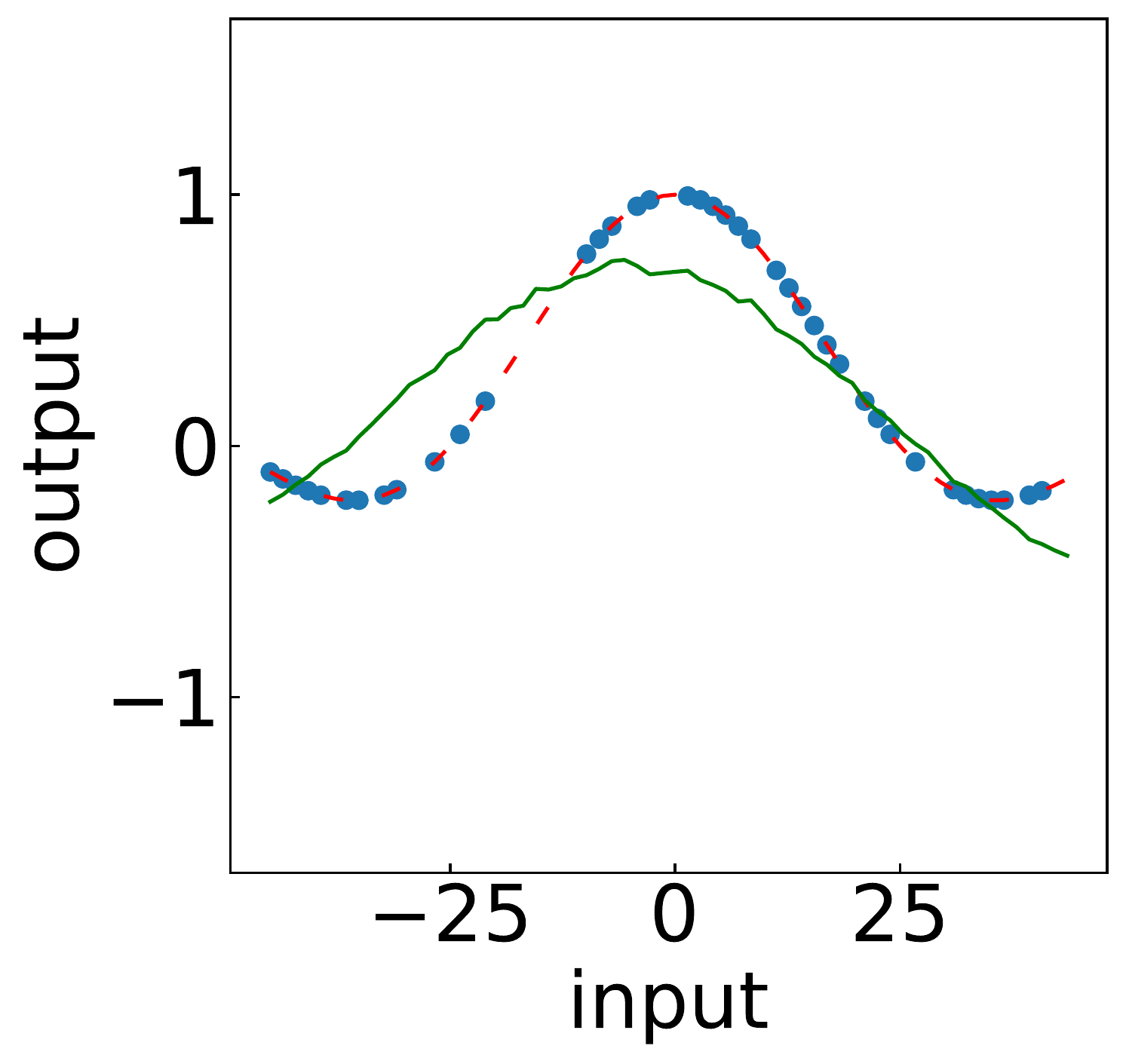}\vspace{-3truemm}
  \subcaption{N = 2}\label{sinc2}
 \end{minipage}\\\vspace{1truemm}
 \begin{minipage}[b]{0.48\linewidth}
  \centering
  \includegraphics[width=37mm]
  {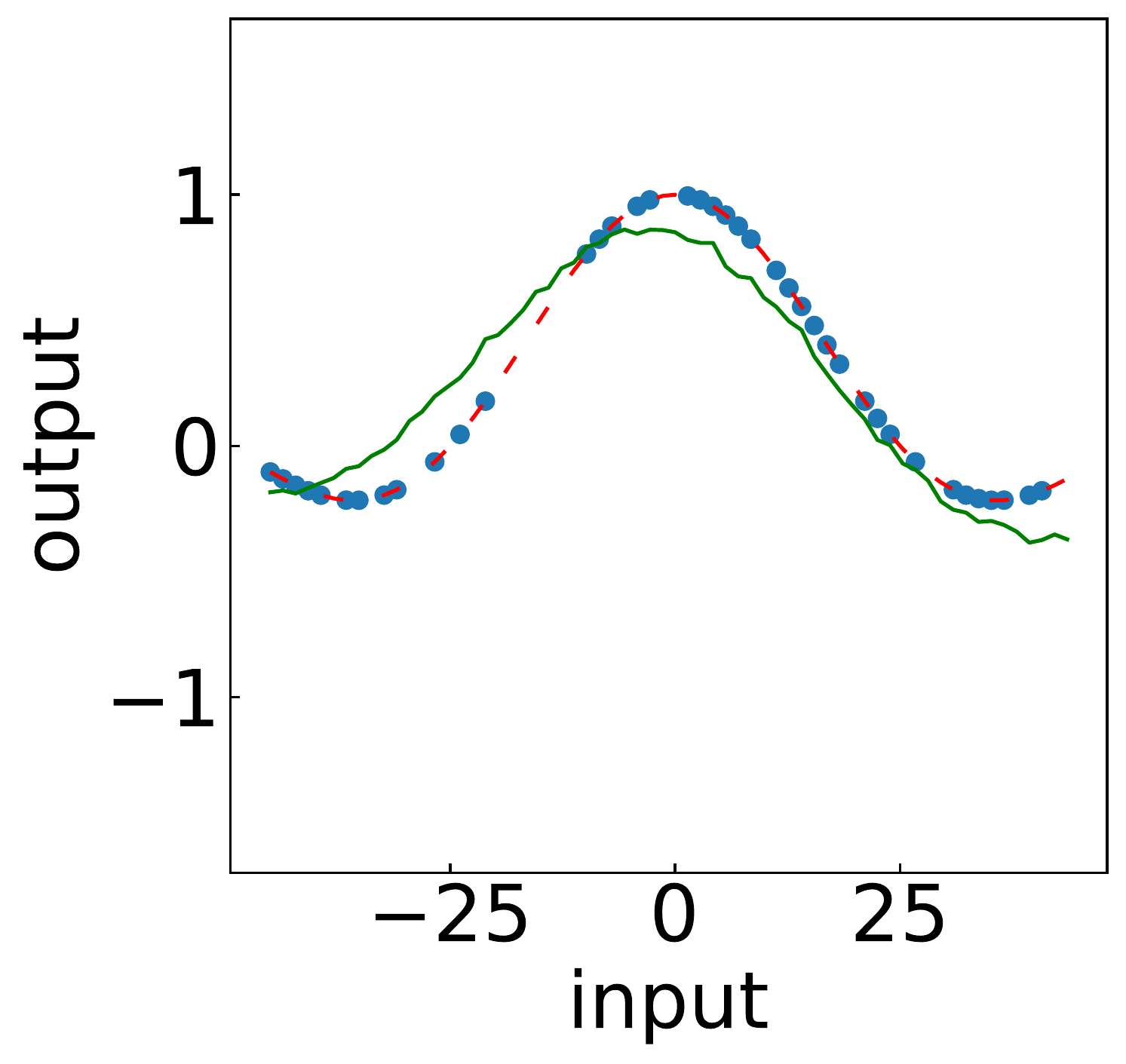}\vspace{-3truemm}
  \subcaption{N = 3}\label{sinc3}
 \end{minipage}
 \begin{minipage}[b]{0.48\linewidth}
  \centering
  \includegraphics[width=37mm]
  {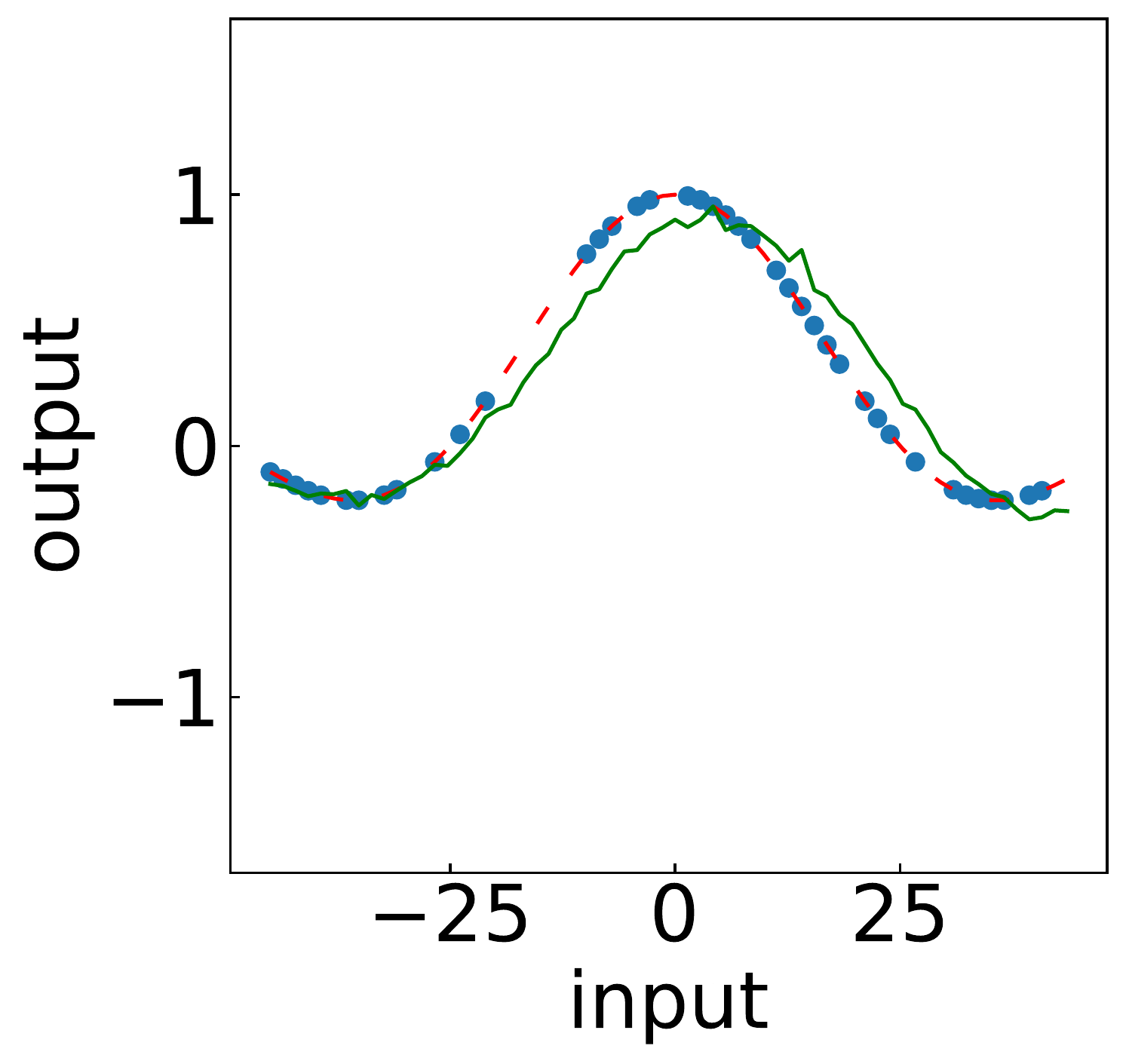}\vspace{-3truemm}
  \subcaption{N = 4}\label{sinc4}
 \end{minipage}\\\vspace{1truemm}
 \begin{minipage}[b]{0.48\linewidth}
  \centering
  \includegraphics[width=37mm]
  {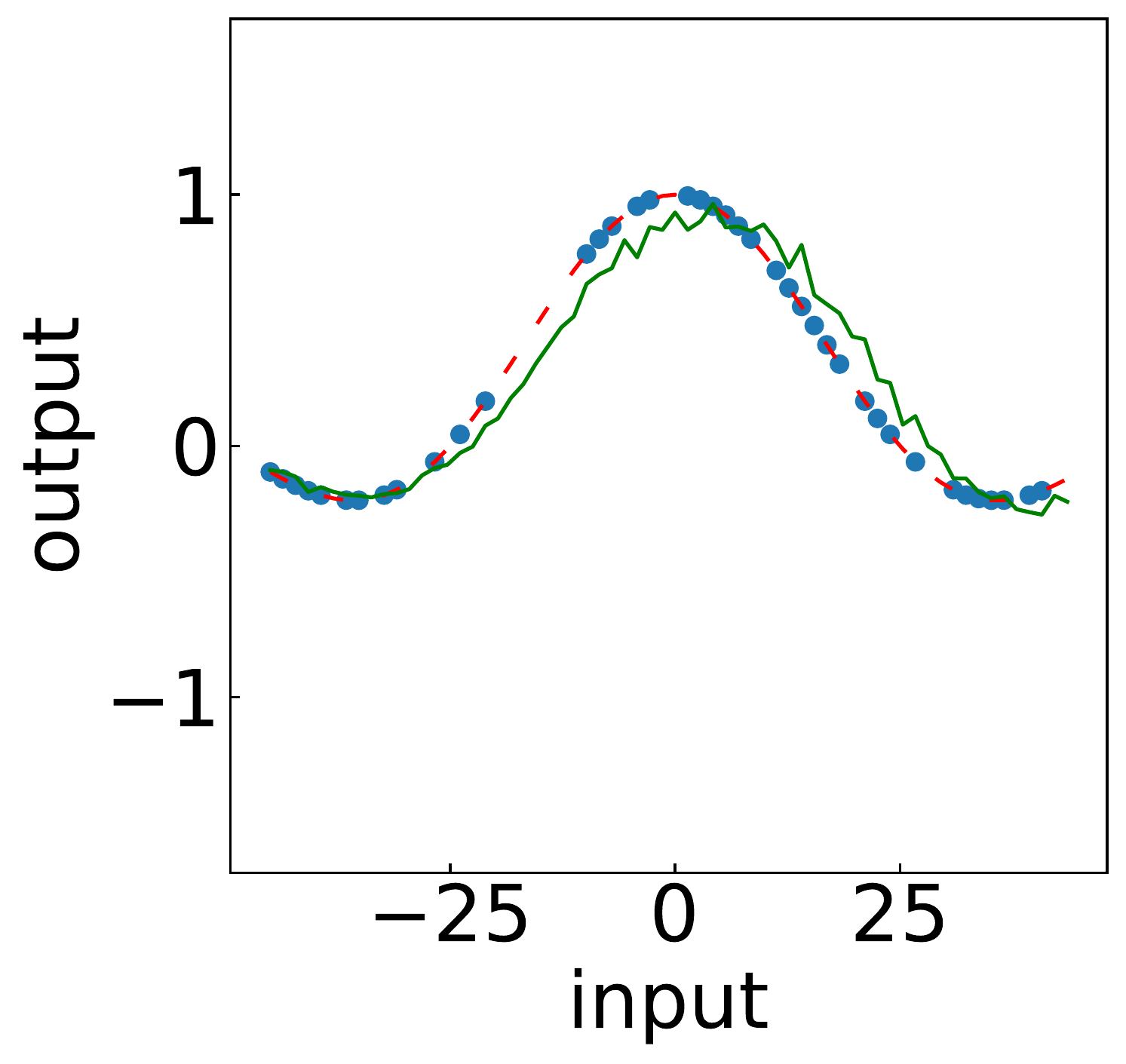}\vspace{-3truemm}
  \subcaption{N = 5}\label{sinc5}
 \end{minipage}
  \begin{minipage}[b]{0.48\linewidth}
  \centering
  \includegraphics[width=37mm]
  {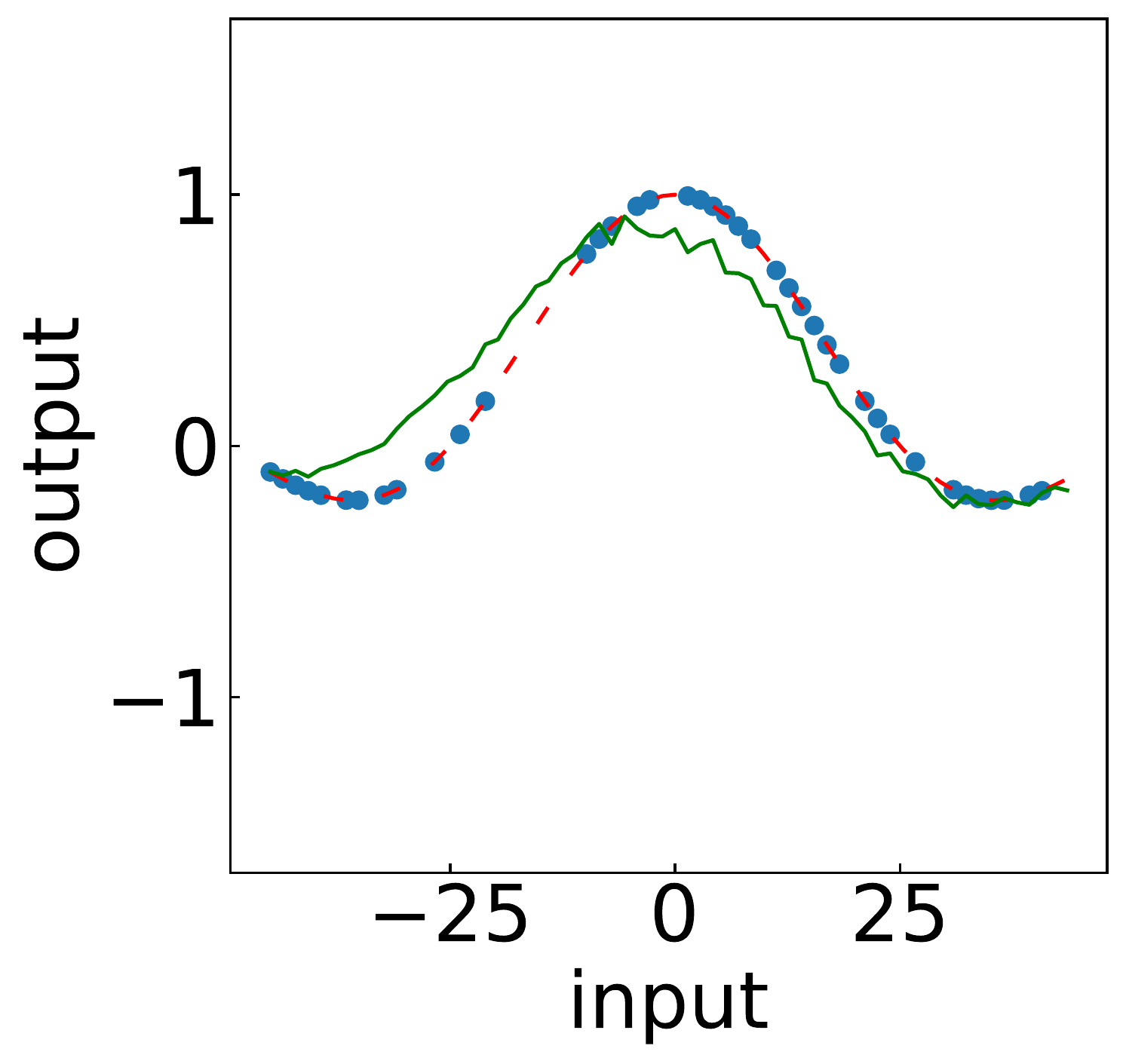}\vspace{-3truemm}
  \subcaption{N = 6}\label{sinc6}
 \end{minipage}
 \end{minipage}\hspace{0.01\linewidth}
 \begin{minipage}[b]{0.48\linewidth}
 \begin{minipage}[b]{0.48\linewidth}
  \centering
  \includegraphics[width=37mm]
  {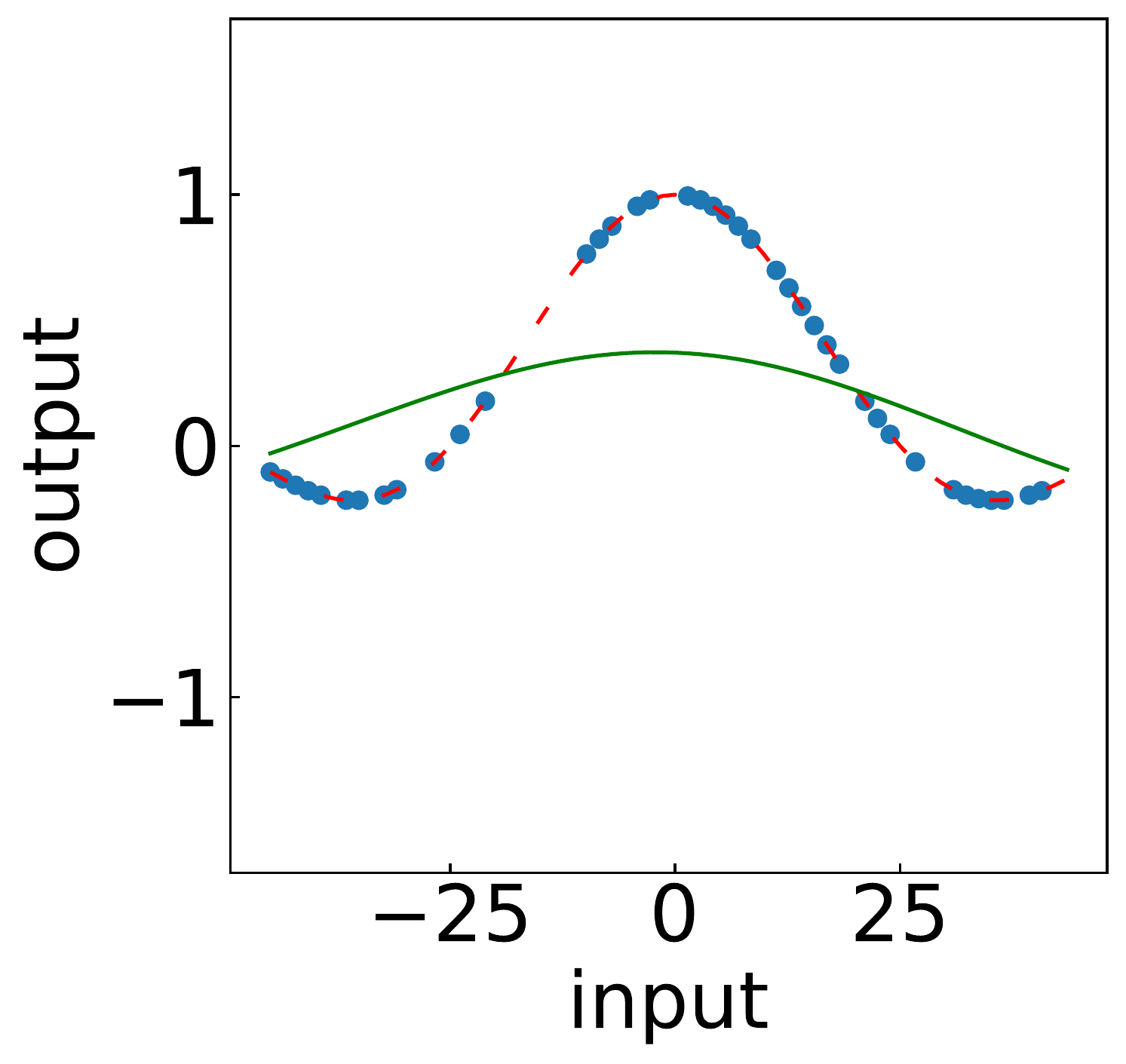}\vspace{-3truemm}
  \subcaption{$\tau=0.02$}\label{fig:sim_sinc1}
 \end{minipage}
 \begin{minipage}[b]{0.48\linewidth}
  \centering
  \includegraphics[width=37mm]
  {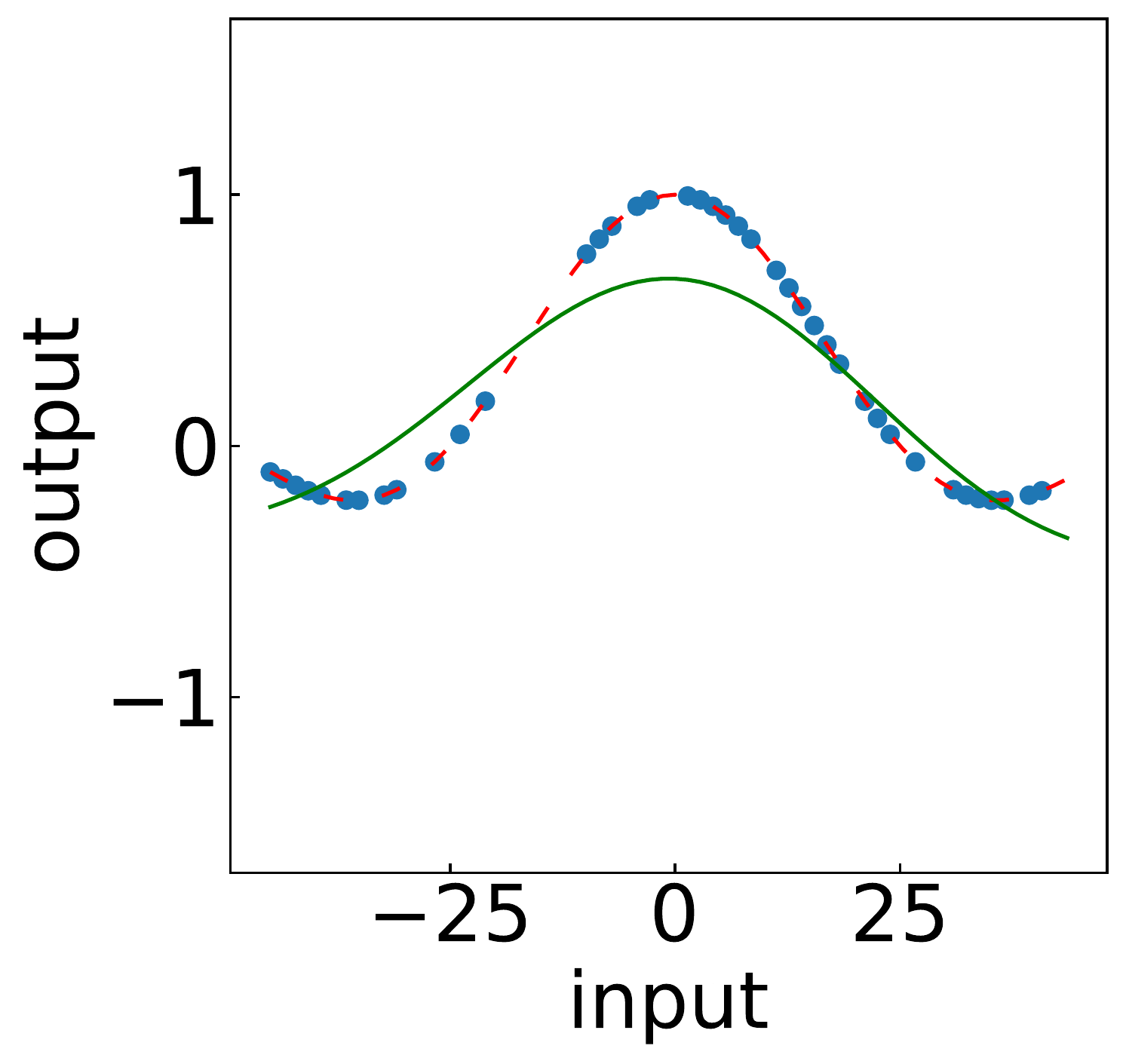}\vspace{-3truemm}
  \subcaption{$\tau=0.04$}\label{fig:sim_sinc2}
 \end{minipage}\\\vspace{1truemm}
 \begin{minipage}[b]{0.48\linewidth}
  \centering
  \includegraphics[width=37mm]
  {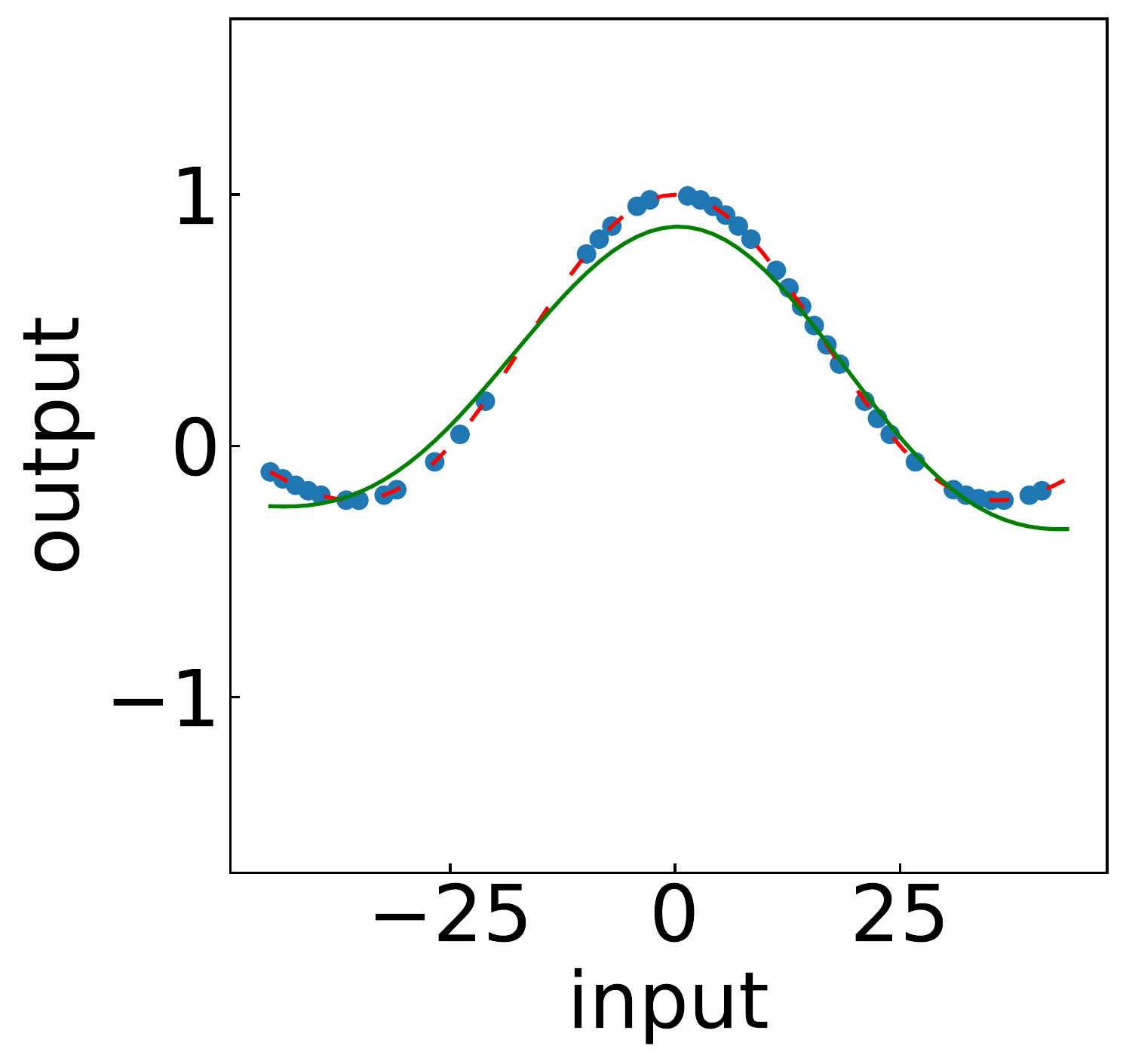}\vspace{-3truemm}
  \subcaption{$\tau=0.06$}\label{fig:sim_sinc3}
 \end{minipage}
 \begin{minipage}[b]{0.48\linewidth}
  \centering
  \includegraphics[width=37mm]
  {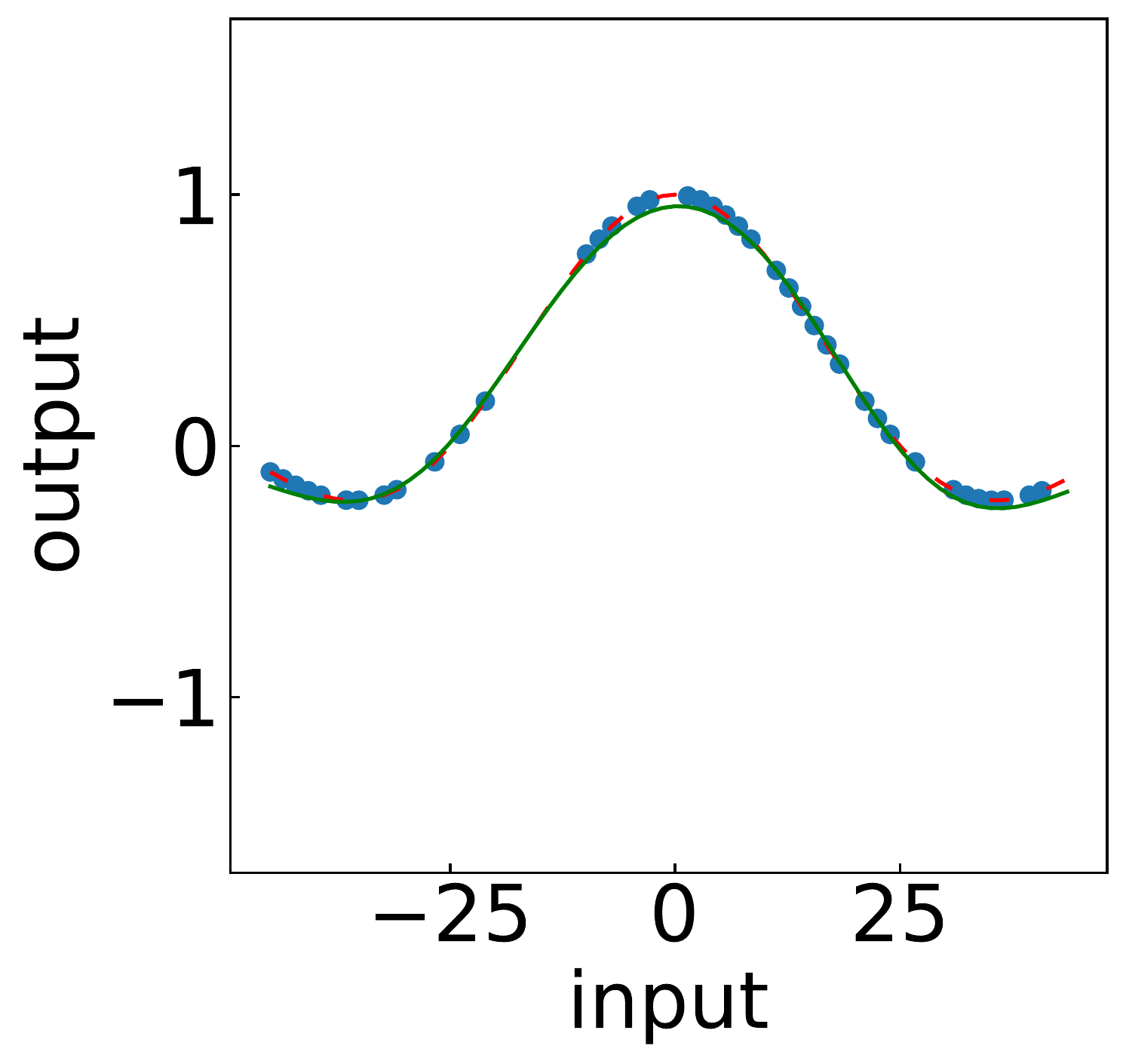}\vspace{-3truemm}
  \subcaption{$\tau=0.08$}\label{fig:sim_sinc4}
 \end{minipage}\\\vspace{1truemm}
 \begin{minipage}[b]{0.48\linewidth}
  \centering
  \includegraphics[width=37mm]
  {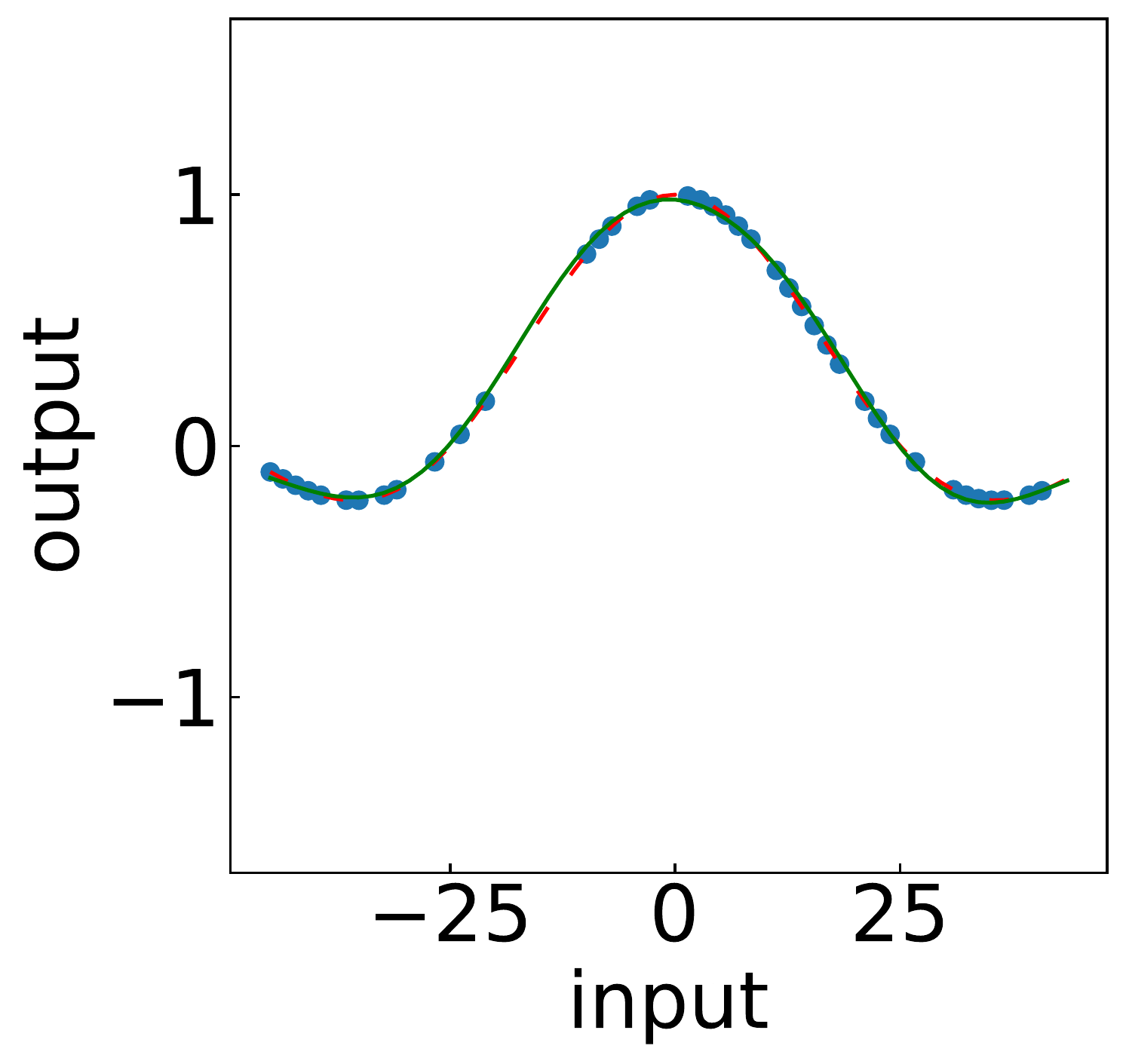}\vspace{-3truemm}
  \subcaption{$\tau=0.10$}\label{fig:sim_sinc5}
 \end{minipage}
 \begin{minipage}[b]{0.48\linewidth}
  \centering
  \includegraphics[width=37mm]
  {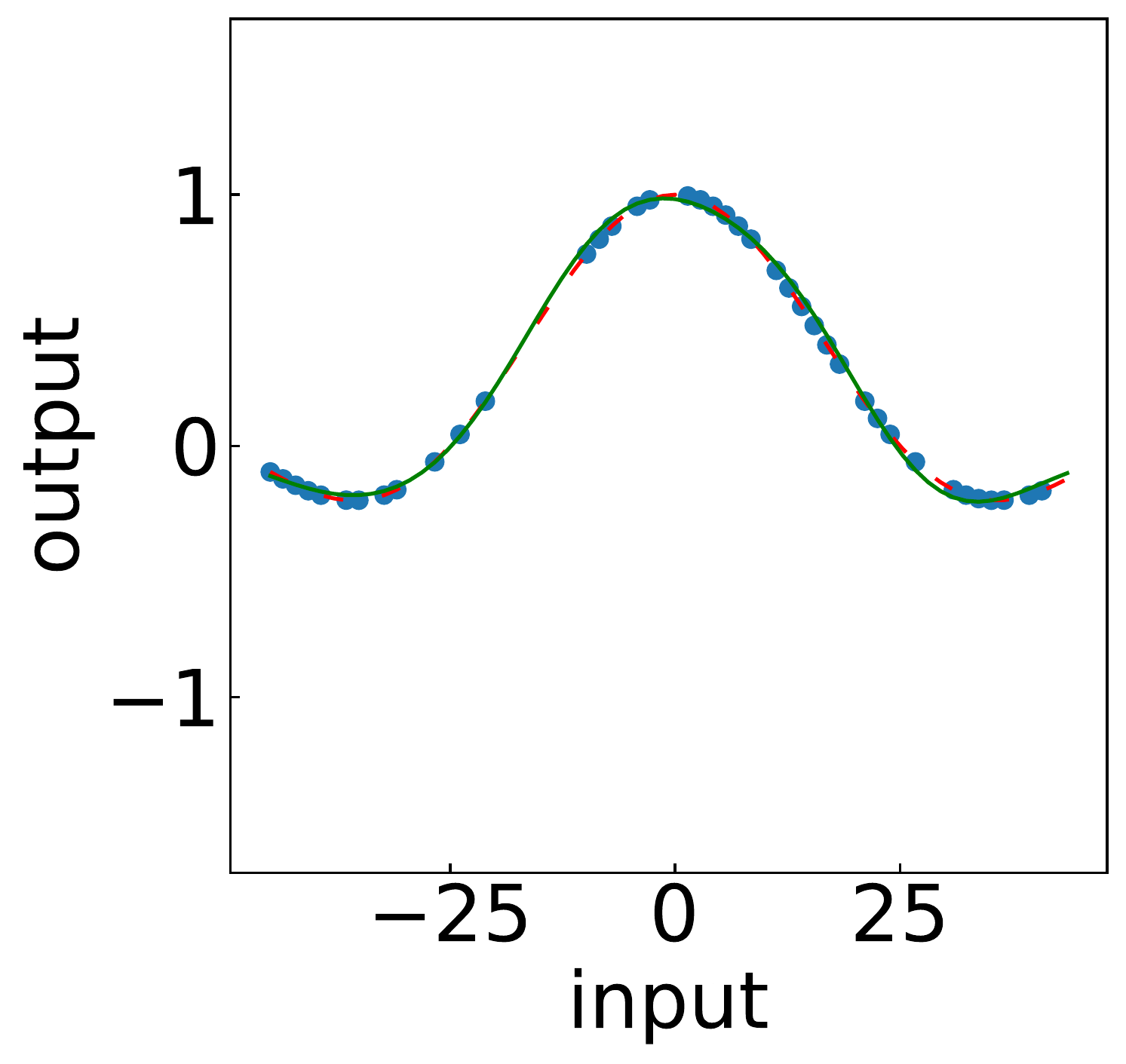}\vspace{-3truemm}
  \subcaption{$\tau=0.12$}\label{fig:sim_sinc6}
 \end{minipage}
 \end{minipage}
 \caption{Demonstration of one-dimensional regression task of $y=\frac{\sin (2\pi x/50)}{2\pi x/50}$ performed with (a)-(f) the NMR kernel and (g)-(l) the numerically simulated kernel. Notations follow those of  Fig.~\ref{fig:sin}. }\label{fig:sinc}
\end{figure*}

We show the results for the regression task of $\mathrm{sinc}$ function performed with the experimental NMR kernel and that of numerical simulations in Fig. \ref{fig:sinc}.

\section{Kernel from the numerical simulations}
\begin{figure}
    \centering
    \includegraphics[width = 0.8\linewidth]{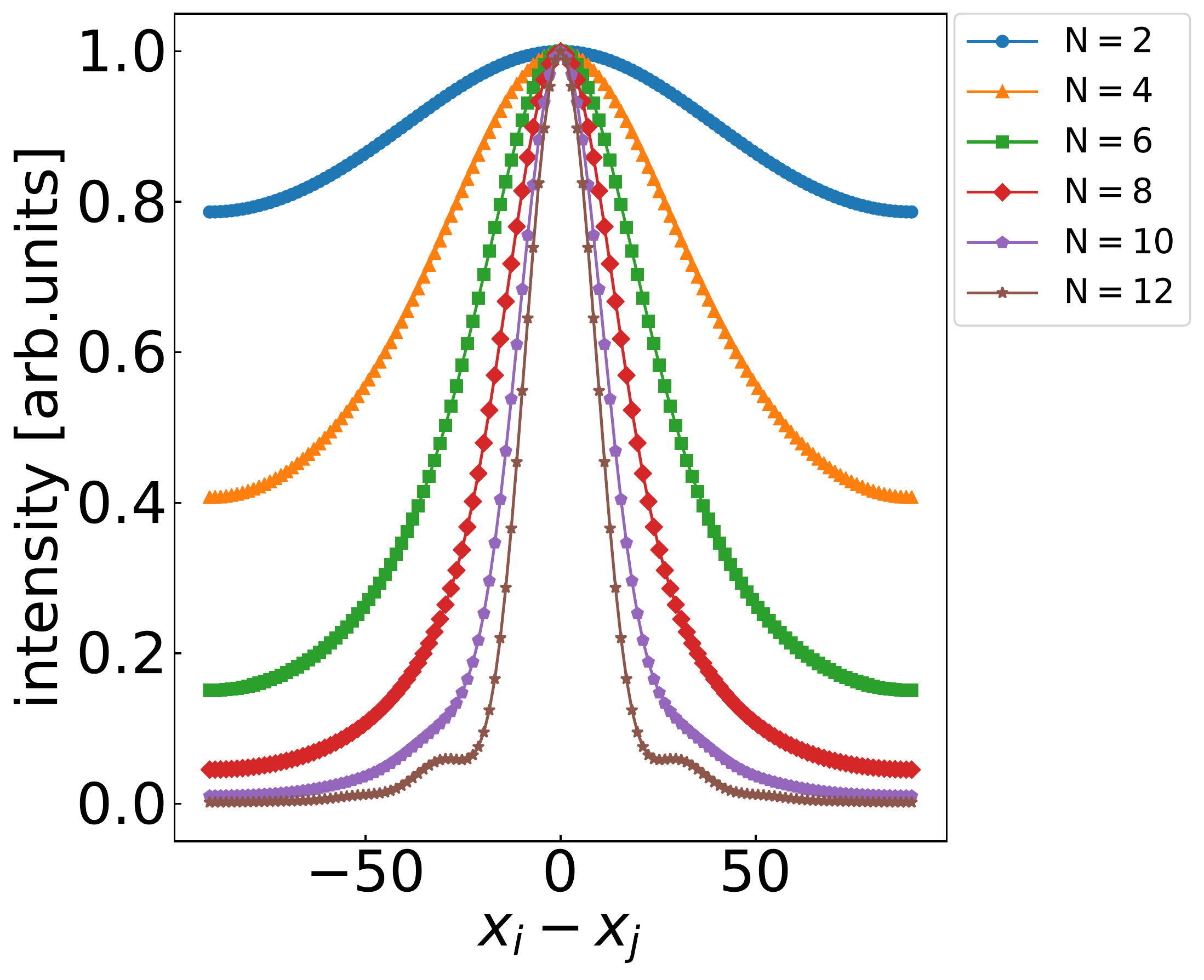}
    \caption{Kernel from the numerical simulations. Here, $\tau = 0.01 N$.}
    \label{fig:sim1DKernel}
\end{figure}

The kernel computed from numerical simulations for one-dimensional input is shown as Fig.~\ref{fig:sim1DKernel}.
It is shown as a function of $x-x'\in\left[-\frac{\pi}{2},\frac{\pi}{2}\right]$.
We can observe the similar features as the experimental one, such as the sharpening of the kernel with increasing evolution time.

\section{Classification tasks}
\subsection{Hinge loss of trained classification model}
We define the hinge loss by, $\frac{1}{N}\sum_{i}\max\{1-\lambda_iy_i,0\}$,
where $N$, $\lambda_i$ and $y_i$ are the number of data, the model output and the teacher for the $i$-th data, respectively.
The hinge loss for each result is computed with the training dataset and shown in Tabs. \ref{tab:HL_exp} and \ref{tab:HL_sim}.
As mentioned in the main text, for the moon dataset, the $N=1$ NMR kernel produced a singular matrix and the result is unreliable.

\begin{table}[H]
    \centering
    \caption{Hinge loss of the trained model with the experimental NMR kernel.}
    \label{tab:HL_exp}
    \begin{tabular}{cccc}
    $N$ & $1$ & $2$ & $3$ \\\hline
    circle & $0.043$ & $0.27$ & $0.15$ \\
    moon & $1.7\times10^{19}$ & $1.0$ & $0.23$ \\
    \end{tabular}
\end{table}

\begin{table}[H]
    \centering
    \caption{Hinge loss of the trained model with the simulated kernel.}
    \label{tab:HL_sim}
    \begin{tabular}{cccc}
    $\tau$  & $0.03$ & $0.06$ & $0.09$ \\\hline
    circle & $0.85$ & $0.17$ & $0.22$ \\
    moon & $0.40$ & $0.19$ & $0.27$ \\
    \end{tabular}
\end{table}

\subsection{Kernel for two-dimensional data}
Let $\bm{x}=(x_1,x_2)$, $\bm{x}'=(x_1',x_2')$ be two data points with which we wish to evaluate the kernel $k(\bm{x},\bm{x}') = k(\{x_1,x_2\},\{x_1',x_2'\})$.
Since our encoding of the data is performed by the unitary defined by Eq. (\ref{eq:uforkernel}) of the main text, the kernel satisfies the equality: $k(\{x_1,x_2\},\{x_1',x_2'\}) = k(\{x_1-x_2',x_2-x_2'\},\{x_1'-x_2',0\})$.
With this in mind, we define $P_1=x_1-x_2'$, $P_2 = x_2-x_2'$, $P_3 = x_1'-x_2'$.
The value of the experimental and simulated kernel is sliced by the value of $P_3$ and shown in Fig. \ref{fig:2Dkernel1}-\ref{fig:2Dkernel3} and Figs. \ref{fig:2Dkernel_sim3}-\ref{fig:2Dkernel_sim9}.

\newcommand*{\getlength}[1]{\strip@pt#1}
\newcount\K 
\newdimen\P 
\def\myforNone#1{
  \P=-90pt           
  \K=0 \loop\ifnum\K<17
  \begin{minipage}[b]{0.24\linewidth}
        \centering
        \includegraphics[width=\linewidth]
        {figure/kernel2D/exp/1cycle/mesh_P3_\number\K_th.png}\vspace{-5pt}
        $P_3$=\getlength{\P}\vspace{4pt}
　\end{minipage}
  \advance\P by11.25pt
  \advance\K by1\repeat 
  }
\begin{figure*}
    \myforNone{}
    \caption{Experimental NMR kernel with $N=1$ for two-dimensional classification task. See text for definitions of $P_1, P_2, P_3$.}
    \label{fig:2Dkernel1}
\end{figure*}

\def\myforNtwo#1{
  \P=-90pt           
  \K=0 \loop\ifnum\K<17
  \begin{minipage}[b]{0.24\linewidth}
        \centering
        \includegraphics[width=\linewidth]
        {figure/kernel2D/exp/2cycle/mesh_P3_\number\K_th.png}\vspace{-5pt}
        $P_3$=\getlength{\P}\vspace{4pt}
　\end{minipage}
  \advance\P by11.25pt
  \advance\K by1\repeat 
  }
\begin{figure*}
    \myforNtwo{}
    \caption{Experimental NMR kernel with $N=2$ for two-dimensional classification task. See text for definitions of $P_1, P_2, P_3$.}
    \label{fig:2Dkernel2}
\end{figure*}

\def\myforNthree#1{
  \P=-90pt           
  \K=0 \loop\ifnum\K<17
  \begin{minipage}[b]{0.24\linewidth}
        \centering
        \includegraphics[width=\linewidth]
        {figure/kernel2D/exp/3cycle/mesh_P3_\number\K_th.png}\vspace{-5pt}
        $P_3$=\getlength{\P}\vspace{4pt}
　\end{minipage}
  \advance\P by11.25pt
  \advance\K by1\repeat 
  }
\begin{figure*}
    \myforNthree{}
    \caption{Experimental NMR kernel with $N=3$ for two-dimensional classification task. See text for definitions of $P_1, P_2, P_3$.}
    \label{fig:2Dkernel3}
\end{figure*}

\def\myfortauthreesim#1{
  \P=-90pt           
  \K=0 \loop\ifnum\K<16
  \begin{minipage}[b]{0.24\linewidth}
        \centering
        \includegraphics[width=\linewidth]
        {figure/kernel2D/simulation/sim3/mesh_2_\number\K_th.pdf}\vspace{-5pt}
        $P_3$=\getlength{\P}\vspace{4pt}
　\end{minipage}
  \advance\P by11.25pt
  \advance\K by1\repeat 
  }
\begin{figure*}
    \myfortauthreesim{}
    \caption{Numerically simulated kernel with $\tau=0.03$ for two-dimensional classification task. See text for definitions of $P_1, P_2, P_3$.}
    \label{fig:2Dkernel_sim3}
\end{figure*}

\def\myfortausixsim#1{
  \P=-90pt           
  \K=0 \loop\ifnum\K<16
  \begin{minipage}[b]{0.24\linewidth}
        \centering
        \includegraphics[width=\linewidth]
        {figure/kernel2D/simulation/sim6/mesh_6_\number\K_th.pdf}\vspace{-5pt}
        $P_3$=\getlength{\P}\vspace{4pt}
　\end{minipage}
  \advance\P by11.25pt
  \advance\K by1\repeat 
  }
\begin{figure*}
    \myfortausixsim{}
    \caption{Numerically simulated kernel with $\tau=0.06$ for two-dimensional classification task. See text for definitions of $P_1, P_2, P_3$.}
    \label{fig:2Dkernel_sim6}
\end{figure*}

\def\myfortauninesim#1{
  \P=-90pt           
  \K=0 \loop\ifnum\K<16
  \begin{minipage}[b]{0.24\linewidth}
        \centering
        \includegraphics[width=\linewidth]
        {figure/kernel2D/simulation/sim9/mesh_9_\number\K_th.pdf}\vspace{-5pt}
        $P_3$=\getlength{\P}\vspace{4pt}
　\end{minipage}
  \advance\P by11.25pt
  \advance\K by1\repeat 
  }
\begin{figure*}
    \myfortauninesim{}
    \caption{Numerically simulated kernel with $\tau=0.09$ for two-dimensional classification task. See text for definitions of $P_1, P_2, P_3$.}
    \label{fig:2Dkernel_sim9}
\end{figure*}

\end{document}